\documentstyle[11pt,epsf,amssymb]{article}

\makeatletter
\@addtoreset{equation}{section}
\makeatother

\def\>{\rangle} %right angle

\def\<{\langle} %left angle

\def\a{\alpha}
\def\b{\beta}

\def\d{\delta}
\def\e{\epsilon}           % Also, \varepsilon
\def\f{\phi}               %      \varphi
  
\def\g{\gamma}

\def\l{\lambda}

  \def\th{\theta}                  %     \vartheta
\def\r{\rho}                                     %     \varrho
\def\t{\tau}

\def\D{\Delta}

\def\G{\Gamma}

\def\del{\partial}              % overwritten by \nabla

\def\cn{{\cal N}}
\def\co{{\cal O}}
\def\cp{{\cal P}}

\def\dalemb#1#2{{\vbox{\hrule height .#2pt
        \hbox{\vrule width.#2pt height#1pt \kern#1pt
                \vrule width.#2pt}
        \hrule height.#2pt}}}

\def\0{{\sst{(0)}}}
\def\1{{\sst{(1)}}}
\def\2{{\sst{(2)}}}
\def\3{{\sst{(3)}}}
\def\4{{\sst{(4)}}}
\def\5{{\sst{(5)}}}
\def\6{{\sst{(6)}}}
\def\7{{\sst{(7)}}}
\def\8{{\sst{(8)}}}

\def\ep{\epsilon}

\def\half{{\textstyle{1\over2}}}
\def\qu{{\textstyle{1\over 4}}}

\let\a=\alpha \let\b=\beta \let\g=\gamma \let\d=\delta \let\e=\epsilon
    
\let\l=\lambda    \let\r=\rho
 \let\t=\tau  \let\f=\phi  
\let\th = \theta
  \let\D=\Delta  
    
    \let\G=\Gamma
\let\la=\label  
   
\def\nn{\nonumber} \def\bd{\begin{document}} \def\ed{\end{document}}
\def\ds{\documentstyle} \let\fr=\frac \let\bl=\bigl \let\br=\bigr
\let\Br=\Bigr \let\Bl=\Bigl 
\let\bm=\bibitem
\let\na=\nabla
\let\pa=\partial \let\ov=\overline 
\newcommand{\be}{\begin{equation}} 
\newcommand{\ee}{\end{equation}}
\def\ba{\begin{array}}
\def\ea{\end{array}}
\def\ft#1#2{{\textstyle{{\scriptstyle #1}\over {\scriptstyle #2}}}}
\def\fft#1#2{{#1 \over #2}}
\def\del{\partial}
\def\sst#1{{\scriptscriptstyle #1}}
\def\oneone{\rlap 1\mkern4mu{\rm l}}
\def\ie{{\it i.e.\ }}
\def\via{{\it via}}
\def\semi{{\ltimes}}
\def\str{{\rm str}}
\def\Dm{{{D_{\sst{max}}}}}

\newcommand{\hsp}{\hspace{0.2cm}}

\newcommand{\ho}[1]{$\, ^{#1}$}
\newcommand{\hoch}[1]{$\, ^{#1}$}
\newcommand{\bea}{\begin{eqnarray}} 
\newcommand{\eea}{\end{eqnarray}} 
\newcommand{\ra}{\rightarrow}
\newcommand{\lra}{\longrightarrow}
\newcommand{\Lra}{\Leftrightarrow}
\newcommand{\ap}{\alpha^\prime}
\newcommand{\tde}{\tilde {\epsilon}}
\newcommand{\tr}{{\rm tr} }
\newcommand{\Tr}{{\rm Tr} } 
\newcommand{\NP}{Nucl. Phys. }
\newcommand{\damtp} {\it DAMTP, University of Cambridge, 
CMS, Wilberforce Road, Cambridge. CB3 0WE}
\newcommand{\spin}{{\it Spinoza Institute, University of Utrecht,\\
Postbus 80.195, 3508 TD Utrecht, The Netherlands}\\
{\tt email:M.TaylorRobinson@phys.uu.nl}}
\newcommand{\princeton}{{\it Physics Department, Princeton University,\\ 
Princeton, NJ 08544, USA}\\
{\tt email:kostas@feynman.princeton.edu}}
\newcommand{\auth}{\large\bf{ 
Kostas Skenderis\hoch{\star} and Marika Taylor\hoch{\dagger}}}

\begin{document}

\begin{flushright}
\hfill{\bf hep-th/0204054}\\
\hfill{SPIN-2002/09} \\
\hfill{ITF-2002/15} \\
\hfill{PUPT-1995}
\end{flushright}

\vspace{15pt}

\begin{center}

{\Large \bf Branes in AdS and pp-wave spacetimes}

\vspace{20pt}

\auth

\vspace{15pt}

{\hoch\star \princeton}

\vspace{8pt}

{\hoch\dagger \spin}

\vspace{15pt}

\underline{ABSTRACT}
\end{center}
\noindent

We find half supersymmetric $AdS$ embeddings in $AdS_5 \times S^5$
corresponding to all quarter BPS orthogonal intersections of D3-branes 
with Dp-branes. A particular case is the Karch-Randall embedding
$AdS_4 \times S^2$. We explicitly prove that these embeddings are 
supersymmetric by showing that the kappa symmetry projections
are compatible with half of the target space Killing spinors and
argue that all these cases lead to AdS/dCFT dualities 
involving a CFT with a defect. We also find an asymptotically 
$AdS_4 \times S^2$ embedding that corresponds to a holographic RG-flow
on the defect. We then consider the pp-wave limit of the supersymmetric
$AdS$ embeddings and show how it leads to 
half supersymmetric D-brane embeddings 
in the pp-wave background. We systematically analyze D-brane
embeddings in the pp-wave background along with their supersymmetry.
We construct supersymmetric D-branes wrapped along the 
light-cone using operators
in the dual gauge theory: the open string states are constructed
using defect fields. We also find supersymmetric 
D1 (monopoles) and D3 (giant gravitons) branes that 
wrap only one of the light-cone directions. These
correspond to non-perturbative states in the dual gauge theory.

{\vfill\leftline{}\vfill
\pagebreak
\setcounter{page}{1}

\tableofcontents
\addtocontents{toc}{\protect\setcounter{tocdepth}{2}}

\newpage

\section{Introduction and summary of results}
\noindent

Two of the most interesting recent developments are the extension
of the AdS/CFT duality to conformal field theories with a
defect (AdS/dCFT duality) 
\cite{bachas2,randall,KR,ooguri1, bachas1,lee,ponsot,johanna}, 
and the study of the pp-wave limit of AdS
backgrounds \cite{blau0,blau,BMN,itzhaki,ooguri3,zayas,alishahia,
kim,takayanagi,floratos,billo,cvetic1,gursoy,michelson,chu,cvetic2,
dabholkar,BMN2,jer,Lee:2002cu,Das:2002cw,Lu:2002kw,Kiritsis:2002kz,
Kumar:2002ps,Leigh:2002pt,Bak:2002rq}. 
In the former case one adds additional structure
on both sides of the duality: a D-brane in the bulk 
and a defect in the boundary theory. 
The theory on the defect captures holographically the physics of the D-brane
in the bulk, and the interactions between the bulk and the D-brane
modes are encoded in the couplings between the boundary 
and the defect fields.  

The pp-wave limit of an AdS background 
is a special case of a Penrose limit 
of a gravitational background \cite{penrose,guven,blau0,blau}. 
The particular interest in 
this Penrose limit stems from the fact that one can combine the limit 
with the AdS/CFT duality to obtain a relation between string 
theory on a pp-wave background and a specific limit of 
the dual conformal field theory \cite{BMN}. It turns out that 
string theory on the pp-wave background is exactly 
solvable \cite{metsaev,russo,mt} and this raises the possibility 
of understanding quantitatively holography for a background 
which is very close to flat space. 

In this paper we give support to and propose a whole host
of new AdS/dCFT dualities by studying D-brane embeddings
in $AdS_5 \times S^5$ and analyzing their supersymmetry. We 
show that the supersymmetric $AdS$-embeddings that we find 
lead to supersymmetric D-brane embeddings
on pp-wave backgrounds. Utilizing the AdS/dCFT duality,
we construct  the light-cone D-brane states 
using gauge theory operators for all D-branes.

In the AdS/dCFT duality proposed by Karch and Randall \cite{KR}, 
one considers a D5 brane wrapping an $AdS_4 \times S^2$
submanifold of $AdS_5 \times S^5$. This configuration may be considered
as the near-horizon limit of a certain D3-D5 system, and the AdS/CFT 
duality is considered to act twice: both in the bulk and on the worldvolume.
In the limit discussed in \cite{ooguri1}, 
the bulk description is in terms of supergravity coupled 
to a probe D5-brane. The dual theory is $\cn=4$ SYM theory coupled to 
a three dimensional defect. The defect theory may be associated with
the boundary of $AdS_4$, and as such it should be a conformal 
field theory. 

An important issue is whether the $AdS_4 \times S^2$ embedding 
is stable and supersymmetric. 
In the probe computation of \cite{KR} it was found that the 
D-brane configuration is sitting in the
maximum of the potential and that there is a tachyonic mode.
The mass of the tachyonic mode was shown to be
above the Breitenlohner-Freedman bound \cite{BF}, signaling stability.
Further evidence for stability and 
supersymmetry of the configuration was provided in \cite{ooguri1} 
where it was shown that one can fit the KK bosonic fields of the
$S^2$-reduction of the worldvolume theory in appropriate supermultiplets.
We prove in the paper that the configuration is supersymmetric,
thus dispelling any doubts about the stability of the system.
In particular, we show that the $AdS_4 \times S^2$ configuration 
with or without flux on the $S^2$ preserves 16 supercharges by showing that 
the worldvolume kappa symmetry projection is compatible with 16 of the 
32 target space Killing spinors. 

In the AdS/CFT correspondence one may study holographic RG-flows
by considering solutions that preserve $d$-dimensional Poincare
invariance and are asymptotically $AdS_{d+1}$. The 
asymptotic behavior of the bulk field shows whether the solution
describes an operator deformation or a new (non-conformal) vacuum of 
the CFT. One expects a similar story in the AdS/dCFT duality.
In particular, one may deform the $\cn=4$ SYM and/or the defect CFT.
In the former case, one should study D5 embeddings on the 
asymptotically $AdS$ solution describing the RG-flow. The 
embeddings are expected to only be asymptotically $AdS$,
signaling an induced RG-flow on the defect theory. 
This is an interesting subject which, however, we will not pursue 
further in this paper.  

The second possibility is to only 
deform the defect CFT. Within the approximations used in 
this paper, the D5-brane theory does not backreact on the 
bulk. This means that the boundary theory remains conformal,
but the defect theory runs. This situation should be
described by an asymptotically $AdS_4 \times S^2$ embedding
in $AdS_5 \times S^5$. We indeed find such an embedding,
where only the $AdS_4$ part of the solution is deformed.
This means that the defect QFT still has the same R-symmetry
as the defect CFT. Furthermore, this embedding completely
breaks supersymmetry. Using the operator-field dictionary 
developed in \cite{ooguri1}, and the asymptotic form of the 
worldvolume fields we show that the RG-flow corresponds 
to a vev deformation of the defect theory. 
The operator that gets the vev is a specific scalar component of 
a $3d$ superfield, so its vev breaks supersymmetry.
Furthermore, it is an R-singlet in accordance 
with the fact that the $S^2$ part of the solution is undeformed.
An interesting feature of this RG-flow is that the theory
develops a mass gap in the infrared. We propose that the 
vev corresponds to giving masses to the 3-5 strings.

We also find solutions where the D5-brane wraps an $S^2$ 
with vanishing volume but non-zero flux. Some of these
solutions are supersymmetric and some not but 
in all cases the solution can be re-interpreted as a D3-brane.
The non-supersymmetric cases correspond to either anti-D3
branes or D3-branes misaligned with the background.
The supersymmetric case correspond to D3 branes aligned 
with the D3-branes creating the background. Next
we point out that $AdS_{m+1} \times S^{n+1}$ embeddings
with general $(m,n)$ in this background satisfy the field 
equations. They are, however, only supersymmetric when $\left| m - n
\right | =2$; this follows from the well-known intersection
rules for intersecting D-brane systems, as we will discuss further below. 

We next consider the pp-wave Penrose limit. 
In the case of $AdS_5 \times S^5$,
the corresponding field theory limit is the large $N$ and $J$ limit,
where $J$ is R-charge associated with a specific $SO(2)$ subgroup
of the R-symmetry group \cite{BMN}. In this limit, the authors 
of \cite{BMN} (BMN) identified certain field theory operators
with closed string states in the pp-wave background. 
In the duality of \cite{KR}, the bulk theory involves the degrees of freedom
of a single D5 brane, and these are encoded holographically in the 
defect CFT. This suggests that the corresponding pp-wave 
limit will capture the degrees of freedom of the D5 brane as well,
and that the field theory operators that correspond to open
string states on the pp-wave background are defect 
operators. This observation was also made independently 
in a nice paper \cite{Lee:2002cu} that appeared while this
work was finalized. 

In the pp-wave spacetime four of the transverse coordinates,
$y^a$, originate from $AdS_5$ and the other four, $z^a$, from $S^5$. 
We denote this splitting of transverse coordinates as $(4,4)$,
or more explicitly as $(y^a;z^a)$.
We find that the Penrose limit of the $AdS_4 \times S^2$ embeddings in 
$AdS_5 \times S^5$ are D5-brane embeddings in the 
pp-wave background which are longitudinal to the 
light-cone (provided that the original brane wraps the boosted circle).
The induced geometry is that of a
pp-wave localized in certain coordinates. In particular,
the $AdS_4 \times S^2$ embedding with zero flux over the $S^2$
yields a pp-wave localized at the origin of $(1,3)$, i.e. $(0;0,0,0)$.
Turning on a flux $q$ on the $S^2$ has the effect of shifting the 
position of the pp-wave in the $AdS$-direction by an amount equal to the flux,
i.e. the D5 brane is now located at $(\pm q; 0, 0, 0)$. 

By a direct computation we show that the solution with or 
without flux preserves half of the supersymmetry, in agreement with the 
fact that the original brane preserved 16 supercharges. (The Penrose
limit always at least preserves supersymmetry.) 
One can also have D5-brane embeddings located at arbitrary
constant positions in the $(1,3)$ directions. These branes, however, 
preserve only one quarter of the 
supersymmetry. This is a generic phenomenon: brane embeddings
with constant transverse scalars generically preserve 1/4 of 
supersymmetry. In some cases, however, the supersymmetry is enhanced
by a factor of two when the brane is located at special points
which indicates that the D-brane configuration originates 
from a 1/2 supersymmetric $AdS$-embedding in $AdS_5 \times S^5$.

Motivated by these results we then investigate systematically 
D-brane embeddings in the pp-wave background and their supersymmetry.
We restrict ourselves (mostly) to embeddings with constant transverse
scalars and zero flux. In some sense these are the ``elementary''
embeddings and more complicated ones may be considered as 
``superpositions'' or ``excitations'' of the elementary embeddings. 
The branes can be divided into longitudinal
branes where the light-cone coordinates are on the worldvolume,
instantonic branes, where the lightcone coordinates are in the
transverse coordinates and branes with one lightcone 
coordinate along the worldvolume and the other 
transverse to it. In light-cone open string theory, only
the first branes are visible and in light-cone closed string theory
one can only construct boundary states for the instantonic 
ones. Although the last branes are not visible in the light-cone gauge,
they should be present in covariant gauges. Except for 
a few exceptional cases (to be discussed in the main text),
we find solutions for all possible splittings of (constant) transverse 
and longitudinal coordinates. 
 
Only branes with specific coordinate splitting preserve supersymmetry.
When they are sitting at arbitrary (constant) positions 
of the transverse space
they preserve one quarter of supersymmetry, but they
preserve 16 supercharges when located at the origin 
of the transverse space. As discussed in the 
previous paragraph, these cases indicate that the solution
originates from the 1/2 supersymmetric $AdS$ embedding
on $AdS_5 \times S^5$. The latter embeddings, in turn, are 
expected to originate from the near-horizon limit of 
supersymmetric intersections of D3 branes with other Dp-branes.
Indeed, inspection of our results confirms this picture and
can be summarized in the following Table 1.

\begin{table}[h]
\begin{center}
\vspace{.2cm}
\begin{tabular}{|c|c|c|c|}
\hline 
Brane & $ND=4$ intersections & Embedding & Longitudinal
\\ 
\hline \hline
D1 & $(0| D1 \perp D3)$ & $AdS_2$              &   -           
\\ \hline
D3 & $(1| D3 \perp D3)$ & $AdS_3 \times S^1$   & $(+,-,2,0)$   
\\ \hline
D5 & $(2| D5 \perp D3)$ & $AdS_4 \times S^2$   & $(+,-,3,1)$   
\\ \hline
D7 & $(3| D7 \perp D3)$ & $AdS_5 \times S^3$   & $(+,-,4,2)$   
\\ \hline
\hline
 & $ND=8$ intersections & &  \\
\hline \hline
D5 & $(0| D5 \perp D3)$ & $AdS_2 \times S^4$   & $(+,-,1,3)$   
\\ \hline
D7 & $(1| D7 \perp D3)$ & $AdS_3 \times S^5$   & $(+,-,2,4)$   
\\ \hline 
\end{tabular}
\caption{{\it Half supersymmetric branes in pp wave backgrounds
and their relation to brane intersections and half supersymmetric 
branes in $AdS_5 \times S^5$. The notation $(+,-,m,n)$ indicates 
a brane wrapping the light-cone coordinates, $m$ coordinates
that used to be $AdS_5$ coordinates before the Penrose
limit, and $n$ coordinates that used to be $S^5$ coordinates.}
}
\end{center}
\end{table}
In the first column of Table 1 we indicate the D-brane  
embedding. The second column lists the supersymmetric intersection 
of the D-brane under consideration with the background D3-branes. The 
top part  of the table contains supersymmetric intersections where 
the number of Neumann-Dirichlet (ND) boundary conditions is equal 
to four: such intersections
are also sometimes called standard intersections. The bottom
part of the table contains supersymmetric intersections with the number 
of ND boundary conditions equal to eight (non-standard intersections).
The notation $(r|Dp\perp D3)$ means that a $Dp$ brane 
intersects orthogonally a $D3$ brane over an $r$-brane.
The third column indicates the expected half supersymmetric D-brane
embeddings in $AdS_5 \times S^5$.  These embeddings follow 
from the second column by considering the near-horizon limit of the 
D3-branes and taking the $Dp$ brane to 
extend in the radial direction of the D3 brane and have the remaining 
$(p-r-1)$ worldvolume directions wrapped on an $S^{p-r-1}$ sphere in $S^5$.
The fourth column lists 
the half supersymmetric longitudinal Dp-branes that we find 
in this paper, for which we use the notation $(+,-,m,n)$ to denote
the worldvolume directions. Note that the $AdS_2$ embedding necessarily 
leads to a pp-wave D-brane wrapping the $x^-$ light-cone direction, and 
thus decouples in the pp-wave limit. There is a supersymmetric
D1-$(+,-,0,0)$ brane but it preserves only a quarter of supersymmetry
even when located at the origin of transverse space.
We also find that $(+,-,0,4)$ and $(-,+,4,0)$ D5-brane embeddings which carry
a specific worldvolume flux are quarter supersymmetric irrespectively
of their location.
 
Instantonic branes may be obtained from 
the longitudinal ones by formal T-duality along the light-cone
directions. The results so obtained agree with the results
obtained in \cite{billo} using the boundary state approach.

The pp-wave background is symmetric under interchange
of $y^a$ by $z^a$. This implies that the supersymmetric branes should 
come in pairs, $(p,q)$ and $(q,p)$ and we indeed find that the results
in Table 1 have this property, except for one entry. We seem
to be missing the the $D3$-$(+,-,0,2)$ brane. Inspection 
of the table suggests that such a brane would have 
come from a $R \times S^3$ embedding, where 
$R$ stands for the time-like direction;
we indeed find such an embedding \footnote{
There is one $ND=8$ intersection that we have not mentioned
so far because of its unusual reality properties: 
$(-1|D3 \perp D3)$. This intersection
leads to an instantonic $D3$-$(1,3)$-brane in the pp-wave limit.}.
 
Branes which lie along only one direction of the lightcone must
be rotating around the boosted circle. Branes which lie along
the $x^{-}$ direction, which would be called $(-,m,n)$ branes
in the above notation, are shown to have degenerate induced metrics
and are not admissible embeddings; they have infinite energy
as measured by the lightcone Hamiltonian. If we take the Penrose
limit of any $AdS_{m+1} \times S^{n+1}$ brane boosting along a 
circle orthogonal to the brane, under the infinite boost the brane
will be mapped to a $(-,m,n)$ brane and hence disappears from the
physical spectrum. 

Branes which lie along the $x^{+}$ direction are rotating at the
speed of light in the direction of the boost. We find a number of
such embeddings, and show explicitly that some of them are 
supersymmetric. In particular, we find 1/2 supersymmetric 
$(+,0,3)$ and $(+,3,0)$ embeddings, and show how they 
related to the BPS giant graviton branes \cite{Len,Itz,My1}
by a Penrose limit.
We also find 1/4 supersymmetric $D1$-branes,
$(+,1,0)$ and $(+,0,1)$, and the corresponding rotating D-strings
in $AdS_5 \times S^5$. We suggest that the dual gauge theory
interpretation of these branes is as $SU(N)$ magnetic 
monopoles\footnote{We thank Erik Verlinde
for a discussion about this point.}.

\bigskip

Having obtained the supersymmetric $AdS$-embeddings given in 
Table 1 one may generalize the arguments of \cite{KR,ooguri1}
to obtain a new set of AdS/dCFT dualities. In each of the 
cases listed one may consider the limit
described in \cite{ooguri1} to obtain a duality between 
the bulk $AdS_5 \times S^5$ with a single Dp-brane
and $d=4$ $\cn=4$ SYM theory interacting with a defect 
theory. The defect theory is that of the
low energy modes of the 3-p and p-3 strings;
operators on the defect are dual to fields on the Dp-brane. 

We have argued above that the supersymmetric $AdS$-embeddings
are mapped to D-brane embeddings in the pp-wave background.
The duality described in the previous paragraph suggests
that the light-cone D-brane states can be constructed
by defect field operators. We indeed find this to be the case.
We propose that the light-cone open string ground state 
corresponds to an operator made of a large
number of $Z$'s (as in the closed string sector of \cite{BMN}) 
together with a bilinear of the massless modes of the 3-p and p-3 strings.
The proposal for the light-cone ground states in the 
$ND=4$ and $ND=8$ cases 
is given in (\ref{openvac}) and (\ref{openvac1}), respectively.
In both cases we find that the light-cone energy 
agrees with the result obtained by a direct computation in
the open string theory on the pp-wave background. 

As already mentioned,
the longitudinal branes always come in pairs, $(+,-,p,q)$ and $(+,-,q,p)$.
The construction of the former brane goes through an $ND{=}4$
system, and the construction of the latter through
an $ND{=}8$ system. Nevertheless, the symmetry of the 
background implies that the two should be equivalent.
In particular, the ground state energy should be the same and
we indeed find both (\ref{openvac}) and (\ref{openvac1})
have the same light-cone energy!
Furthermore, following \cite{BMN} and the recent 
papers \cite{BMN2,Lee:2002cu}, we construct the oscillators by 
inserting $D_i Z$ and $\phi_i$ operators in closed and open string
ground states and by adding phase (closed strings),
cosine (open strings with Neumann boundary conditions)
and sine (open strings with Dirichlet boundary conditions) factors.

The D5 and D7 branes appear twice in Table 1, and thus in the
decoupling limit one obtains two distinct supersymmetric 
D5/7-brane embeddings into $AdS_5 \times S^5$, with the corresponding
dual theories containing different codimension defects. However, as
we said, in the pp-wave limit the symmetry of the background
under the interchange of the $y^a$ and $z^a$ directions
implies that the $(+,-,m,n)$ branes are indistinguishable
from the $(+,-,n,m)$ branes. This means the surviving sectors
of the two dCFTs dual to the D-branes should be equivalent,
even though they contain defects of different dimensions.
We leave the study of this rather novel ``duality'' for future work. 

\bigskip

The organisation of the paper is as follows. In \S\ref{dbf2} we derive
the D-brane field equations in full generality, for use in later 
sections. In \S\ref{ads4s2} we find D5-brane embeddings into
the $AdS_5 \times S^5$ background. In \S\ref{kapsusy} we 
use the kappa symmetry projector to determine the 
supersymmetry preserved by these D-brane embeddings. In 
\S\ref{holdual} we discuss the holographic interpretation of
asymptotically $AdS_4 \times S^2$ embeddings in terms of an
RG flow of the defect theory. In \S\ref{penrose} we consider
the Penrose limits of our brane embeddings, and explicitly
verify their supersymmetry in the pp-wave background. 
In \S\ref{general} we discuss more generally brane embeddings
into $AdS_5 \times S^5$ and their supersymmetry. In \S\ref{branes}
we derive brane embeddings in the pp-wave background, considering
branes with two, one and zero directions along the light cone.
In \S\ref{gauD} we present the new AdS/dCFT dualities, and
we construct the D-brane states from gauge theory operators.
Finally, in appendix A we list our conventions and in appendix B
we derive and discuss an alternative form of the kappa symmetry projection
used in \S\ref{kapsusy}.

\section{D-brane field equations} \label{dbf2}
\noindent

In this section we derive the D-brane field equations in full generality.
It is common practice in the literature to substitute an ansatz for 
a D-brane embedding in the D-brane action and then derive field 
equations by varying the functions appearing in the ansatz.
A given ansatz, however, may not be consistent, i.e. the 
(components of the) fields that are set to zero may be sourced
by non-zero terms in the actual field equations. As the 
D-brane equations are non-linear, such potential problems
are certainly an issue. To avoid such pitfalls, we derive the 
field equations in all generality and for all Dp-branes
in this section. The result, given in (\ref{finfe}), is rather 
compact and can be effectively used in actual computations. 
That is, it is straightforward to obtain the equations satisfied
by the functions appearing in a given ansatz by just substituting the ansatz 
in (\ref{finfe}). We carry out a number of 
such computations in subsequent sections.

The worldvolume action for a single $Dp$-brane is given by
\bea
I_{p} &=& I_{DBI} +  I_{WZ}  \label{dbiact} \\
I_{DBI} &=&
=- T_{p} \int_{M} d^{p+1}\xi e^{-\Phi} 
\sqrt{-\det \left (g_{ij} + {\cal{F}}_{ij} \right )}, \nonumber \qquad
I_{WZ} = T_{p} \int_{M} e^{\cal{F}} \wedge C, \nonumber
\eea
with $T_p$ the Dp-brane tension, which henceforth we set to one. 
Here $\xi^i$ are the coordinates of the $(p+1)$-dimensional
worldvolume $M$ which is mapped by worldvolume fields $X^m$ into the target
space which has (string frame) metric $g_{mn}$. This embedding induces
a worldvolume metric $g_{ij} = g_{mn} \del_{i} X^{m} \del_{j} X^{n}$.
The worldvolume also carries an intrinsic abelian gauge field $A$ with field
strength $F$. ${\cal{F}} = F - B$ is the gauge invariant two-form with 
$B_{ij}=\del_i X^m \del_j X^n B_{mn}$ the pullback of the target 
space NS-NS 2-form. Note that we set 
$2 \pi \a' = 1$ in all that follows. The RR $n$-form gauge potentials (pulled 
back to the worldvolume) are collected in 
\be
C = \bigoplus_n C_{(n)},
\ee
and the integration over $M$ automatically selects the proper forms in
this sum. To simplify the notation we will denote target space tensors
and their pullbacks by the same letter. One can distinguish between
the two by their indices: $m,n,p,...$ denote target space indices,
and $i,j,k$ pullbacks. For instance,
\be
A_{ijmn}=\del_i X^p \del_j X^q A_{pqmn}
\ee
where $A_{pqmn}$ denotes some target space tensor.

Now let us derive the equations of motion which follow from (\ref{dbiact}).
It is convenient to treat the variation of the DBI and WZ terms separately
and to introduce the notation
\be
M_{ij} = (\del_{i} X^{m} \del_{j} X^{n} g_{mn} - 
\del_{i} X^{m} \del_{j} X^{n} B_{mn} + F_{ij}).
\ee
Let us define the inverse of $M_{ij}$ such that
\be
M^{ij} M_{jk} = \delta^{i}_{\hspace{2mm} k}.
\ee
Then variation of the DBI term gives
\bea
\delta I_{DBI} &=& - \int d^{p+1}\xi e^{-\Phi} \sqrt{-M} \left (
-\Phi_{,m} \delta X^{m}
+ \half M^{ji} \delta M_{ij} \right ); \\
&=& - \int d^{p+1}\xi e^{-\Phi} \sqrt{-M} \left ( 
 G^{ij} (\del_i \delta X^m)
(\del_{j} X^{n}) g_{mn} + \theta^{ij} (\del_{i} \delta X^m) 
(\del_{j} X^n) B_{mn} \right .\nonumber \\
&& \left .
+ (\half G^{ij} \del_{i} X^{n} \del_{j} X^{p} 
g_{np,m}  
+ \half \theta^{ij} \del_{i} X^{n} \del_{j} X^{p} B_{np,m} - 
\Phi_{,m} ) \delta X^{m} 
- \theta^{ij} (\del_{i} \delta A_{j}) \right ), \nonumber
\eea
where we introduce the notation $G^{ij} \equiv M^{(ij)}$ 
and $\theta^{ij} \equiv M^{[ij]}$ (we symmetrize and antisymmetrize
with unit strength).

The gauge field equation is 
\be
J^{j} = \del_{i} (e^{-\Phi} \sqrt{-M} \theta^{ij}), \label{gfe}
\ee
where $J^{j}\equiv \d I_{WZ}/\delta A_j$ 
is the source current derived from varying 
the Wess-Zumino terms. 
The $X^{m}$ field equation is 
\bea
J_{m} &=& - \del_{i} \left (e^{-\Phi} 
\sqrt{-M} G^{ij} (\del_{j} X^n) g_{mn} \right ) 
- \del_{i} \left (e^{-\Phi} 
\sqrt{-M} \theta^{ij} (\del_{j} X^{n}) B_{mn} \right ) \label{com} \\
&& 
+ \sqrt{-M} \left ( \half (e^{-\Phi} 
G^{ij} \del_{i} X^{n} \del_{j} X^{p} g_{np,m}
+ e^{-\Phi}
\theta^{ij} \del_{i} X^{n} \del_{j} X^{p} B_{np,m}) - e^{-\Phi} 
\Phi_{,m} \right ) \nonumber
\eea
where $J_{m}\equiv \d I_{WZ}/\delta X^m$ 
denotes the contribution from the WZ terms in the action. 
We discuss the WZ contributions below.
 
To rewrite the equation in a natural covariant form
we expand out the derivatives and use 
\bea
\G_{mnp} & \equiv & \half (g_{mn,p} + g_{mp,n} - g_{np,m}) \\
H_{mnp} & \equiv & (B_{mn,p} + B_{np,m} + B_{pm,n}), \nonumber
\eea
where $\G_{mnp}$ is the Levi-Civita connection of the target space
metric and $H_{mnp}$ is the field strength of the
NS-NS two form. Then we can express the field equation as
\bea
J_{m} &=& - e^{-\Phi} \del_{i} (\sqrt{-M} G^{ij}) \del_{j} X^{n} g_{mn} 
- \del_{i} (e^{-\Phi} \sqrt{-M} \theta^{ij}) \del_{j} X^{n} B_{mn} 
\label{jm} \\
&& - e^{-\Phi} \sqrt{-M} M^{ij} \left ( (\del_{i} \del_{j} X^{n}) g_{mn} 
   + \tilde{\Gamma}_{mnp} \del_{i} X^{n} \del_{j} X^{p} 
\right ),
\nonumber \\
&& + e^{-\Phi} \sqrt{-M} \left ( G^{ij} (\del_{i} X^{p} \del_{j} X^{n}) g_{mn} 
\Phi_{,p} - \Phi_{,m} \right ) \nonumber
\eea
where we use symmetry to replace $G^{ij}$ by $M^{ij}$ in the first
term of the second line and we introduce the torsionful connection 
$\tilde{\Gamma} = \G - \half H$. 

The gauge field equation (\ref{gfe}) can be used
to substitute for the second term in the first line of (\ref{jm})
to give
\bea
J_{m} + J^{j} \del_{j} X^{n} B_{mn} &=&  
- e^{-\Phi} \del_{i} (\sqrt{-M} G^{ij}) \del_{j} X^{n} g_{mn} \label{jmm} \\
&& - e^{-\Phi} \sqrt{-M} M^{ij} \left ( (\del_{i} \del_{j} X^{n}) g_{mn} 
   + \tilde{\Gamma}_{mnp} \del_{i} X^{n} \del_{j} X^{p} \right ) 
\nonumber \\
&& + e^{-\Phi} \sqrt{-M} \left ( G^{ij} (\del_{i} X^{p} \del_{j} X^{n}) g_{mn} 
\Phi_{,p} - \Phi_{,m} \right ). \nonumber
\eea
When $F_{ij} = B_{mn} = \Phi = 0$, the equation reduces to 
\be
J^{m}=- \sqrt{-g} g^{ij} {\cal K}^{m}_{ij}
\ee
where 
\be
{\cal K}^{m}_{ij}=\gamma^{k}_{ij} \del_{k} X^{m}
- (\del_{i} \del_{j} X^{m}) - \Gamma^{m}_{np} \del_{i}X^{n} \del_{j} X^{p}
\ee
is the second fundamental form ($\gamma^k_{ij}$ is the Levi-Civita 
connection of the induced worldvolume metric).
If in addition $J_{m} = 0$, the field equation becomes
\be
g^{ij} {\cal K}^{m}_{ij} = 0, \label{kcon}
\ee
that is, the trace of the second fundamental form of the embedding
must be zero. For a flat target space, this condition is well-known;
it is given in \cite{Bachas:1999um}, for example. 

\bigskip

Now let us derive the explicit form for the Wess-Zumino contributions.
It is convenient to expand out the Wess-Zumino terms as
\be
I_{WZ} = \sum_{n \ge 0} \frac{1}{n! (2!)^{n} q!} \int d^{p+1} \xi  
\ep^{i_1..i_{p+1}} \left ( ({\cal{F}})^{n}_{i_1..i_{2n}}
C_{i_{2n+1}..i_{p+1}} \right ),
\ee
where $\ep^{i_1... i_{p+1}}$ is the Levi-Civita tensor (with no
metric factors) and $q = (p+1-2n)$. 
Variation of the action then gives
\bea
\delta I_{WZ} &=& \sum_{n \ge 0} \frac{1}{n! (2!)^{n} q!} \int d^{p+1} \xi  
\ep^{i_1...i_{p+1}} \left [
n (2 \del_{i_1} \delta A_{i_2} - B_{mn,p} \delta X^{p} \del_{i_1} X^{m}
\del_{i_2} X^{n} \right . \nonumber \\
&&
- 2 B_{mn} \del_{i_1} (\delta X^m) \del_{i_2} X^{n})
({\cal{F}})^{n-1}_{i_3..i_{2n}} C_{i_{2n+1}...i_{p+1}}  
\nonumber \\
&& 
+ ({\cal F})^{n}_{i_1..i_{2n}} \left ( ( q C_{m_1..m_q} \del_{i_{2n+1}}
\delta X^{m_1} ... \del_{i_{p+1}} X^{m_q}  \right . \nonumber \\
&& \left .
+ C_{m_1...m_q,m} \delta X^{m} \del_{i_{2n+1}}
X^{m_1} ... \del_{i_{p+1}} X^{m_q} ) \right ].
\eea
This gives the following expression for the gauge field current 
appearing in (\ref{gfe})
\be 
J^{i_1} =   \ep^{i_1..i_{p+1}} \sum_{n \ge 0} \frac{1}{n! (2!)^{n} (q-1)!} 
({\cal F})^{n}_{i_2...i_{2n+1}} 
\bar{F}_{i_{2n+2}... i_{p+1}}. \label{ji1}
\ee
where\footnote{Our convention for the field strengths is
$f_{m_1...m_{q+1}} = (q+1) \del_{[m_1} C_{m_{2}...m_{q+1}]}$.}
\be
\bar{F}_{m_1..m_{q+1}} = f_{m_1..m_{q+1}} - \frac{(q+1)!}{3! (q-2)!}
H_{[m_1..m_3} C_{m_4..m_{q+1}]}.
\label{rrfe}
\ee
In (\ref{ji1}) we must sum over all possible values of $n$: 
in particular this means that for $p \ge 4$ we must include in
the WZ term the dual RR potentials $C_{5}$, $C_{7}$ and $C_{9}$ (in type IIA) 
and $C_{6}$, $C_{8}$ (in type IIB)\footnote{
One defines the dual potentials as follows. The D7-brane
in type IIB couples in (\ref{ji1}) to the (pull back of the) 
7-form field strength $\bar{F}_7$ which according to (\ref{rrfe})
satisfies (reverting to form notation),
\be \label{rrfe2}
\bar{F}_{7} = d C_{6} - H_{3} \wedge C_{4}.
\ee
This should be regarded as the defining equation for $C_{6}$, since
we identify $\bar{F}_{7}$ with the (target space) dual of $\bar{F}_3$
and acting with the exterior derivative on (\ref{rrfe2}) leads
to the usual IIB field equation
$d (\ast \bar{F}_3) = H_{3} \wedge f_5$.
Analogous equations can be derived for the other dual potentials.}. 
If we do not include the
dual potentials there will remain gauge dependent terms in (\ref{ji1})
involving $H_{[i_1..i_3} C_{i_4..i_{q+1}]}$.

From the $X^{m}$ variation we find the expression for the current
appearing in (\ref{com}). The expression simplify when we
consider the combination that enters in (\ref{jmm}), and we obtain
\be \label{jmf}
J_{m} + J^{j} \del_{j} X^{n} B_{mn} = \sum_{n \ge 0} 
\frac{1}{n! (2!)^{n} q!} \ep^{i_1..i_{p+1}}
({\cal F})^{n}_{i_1..i_{2n}} \bar{F}_{m i_{2n+1}... i_{p+1}},
\ee
in which again we must include the dual RR potentials. 

Let us summarise the D-brane field equations
\bea
&&\hspace{-10mm}\sum_{n \ge 0} \frac{1}{n! (2!)^n q!} 
\ep^{i_1..i_{p+1}}
({\cal F})^{n}_{i_1..i_{2n}} \bar{F}_{m i_{2n+1}... i_{p+1}}=
e^{-\Phi} \left( 
\sqrt{-M} \left ( G^{ij} \del_{i} X^{p} \del_{j} X^{n} g_{mn} 
\Phi_{,p} - \Phi_{,m} \right ) - {\cal K}_m \right) \nonumber \\
&& \del_{i} (e^{-\Phi} \sqrt{-M} \theta^{ii_1}) =  
 \ep^{i_1..i_{p+1}} \sum_{n \ge 0} \frac{1}{n! (2!)^n (q-1)!} 
({\cal F})^{n}_{i_2...i_{2n+1}} 
\bar{F}_{i_{2n+2}... i_{p+1}}. \label{finfe}
\eea
where 
\be
{\cal K}_m=-\del_{i} (\sqrt{-M} G^{ij}) \del_{j} X^{n} g_{mn} 
-\sqrt{-M} M^{ij} \left ( (\del_{i} \del_{j} X^{n}) g_{mn} 
   + \tilde{\Gamma}_{mnp} \del_{i} X^{n} \del_{j} X^{p} \right ) 
\ee
The gauge invariance of the field equation imply that 
${\cal K}_m$ is gauge invariant.
Furthermore, when $B_{mn}=0$, ${\cal K}_m$  reduces to the 
trace of the second fundamental form. 
It follows that ${\cal K}_m$ is a gauge invariant generalization
of the latter. 

\section{D5-brane embeddings in $AdS_5 \times S^5$} \label{ads4s2}
\noindent

Let us now specialize to D5-brane embeddings in an $AdS_5 \times
S^5$ background. The background geometry is 
\bea
ds_{10}^2 &=& R^2 \left (\frac{du^2}{u^2} + u^2 (dx \cdot dx)_4 
+ ds^2_{S^5} \right ); \la{z1} \nonumber \\
ds_{5}^2 &=& d\theta_1^2 + \sum_{k=2}^5 \prod_{j=1}^{k-1}  
\sin\theta_j^2 d\theta_k^2; \label{ads5}\\
f_{5} &=& 4 R^4 u^3 du \wedge dx^0 \wedge dx^1 \wedge dx^2 
\wedge dx^3 + 4 R^4 d\Omega_5, \nonumber
\eea
where the curvature is quantised as $R^4 = 4 \pi g N (\a')^2$. Conventions
for the IIB field equations are given in the appendix and
since $g N$ does not play a role here we will set $R=1$ henceforth.
Using (\ref{finfe}) the D5-brane field equations reduce to
\bea
\del_{i} (\sqrt{-M} \theta^{i i_1}) &=& \frac{1}{5!} 
\ep^{i_1 i_2 i_3 i_4 i_5 i_6} f_{i_2 i_3 i_4 i_5 i_6}; \label{krfe} \\
\frac{1}{2! 4!} 
\ep^{i_1 i_2 i_3 i_4 i_5 i_6} F_{i_1 i_2} f_{i_3 i_4 i_5 i_6 m} &=&
- \del_{i} (\sqrt{-M} G^{ij} \del_{j} X^{n} g_{mn}) + 
\half \sqrt{-M} (G^{ij} \del_{i} X^{n} \del_{j} X^{p} g_{np,m}). \nonumber
\eea
where we remind our readers that $f_{i_3 i_4 i_5 i_6 m}$ 
denotes the pullback of $f$ on the first four indices, i.e.
$f_{i_3 i_4 i_5 i_6 m} = \del_{i_3} X^{m_3} \del_{i_4} X^{m_4}
\del_{i_5} X^{m_5} \del_{i_6} X^{m_6} f_{m_3 m_4 m_5 m_6 m}$.
The solution set of these equations describes all possible
(including non-static) embeddings of D5-branes into the target
space. Here we are interested in D5-branes which wrap an $S^2$
in the $S^5$ and whose remaining worldvolume directions
preserve Poincar\'{e} invariance in three directions. Such
embeddings can be found from the following
ansatz: split the embedding coordinates $X^{m}$ into $\{ \xi^{i},
X^{\lambda} (\xi^{i}) \}$, where the worldvolume coordinates are
\be
\xi^{i} = \{ x^0, x^1, x^2, u, \theta_4, \theta_5 \} \label{ans1}
\ee
and the transverse scalars are 
\be
X^{\lambda} = \{ x^3(u) \equiv x(u), \theta_1, \theta_2, \theta_3 \}, 
\label{ans2}
\ee
where for ease of notation we relabel $x^3$ as $x$ and we assume
that the only dependence of the transverse scalars on the worldvolume
coordinates is in $x(u)$. We also switch on a worldvolume flux
\be  \label{flux}
F_{\theta_4 \theta_5} = q \sin \theta_4.
\ee
With this ansatz it is straightforward to calculate all the
quantities appearing in (\ref{krfe}); for example,
\be
\sqrt{-M} = u^2 (1 + u^4 (x')^2)^{\frac{1}{2}} L_{\th_{\a}}
\sin \theta_4, \label{m1}
\ee
where prime denotes the derivative with respect to $u$ and
\be
L_{\th_{\a}} =(\prod_{\a=1}^{3} \sin^4 \theta_{\a} + q^2)^{\frac{1}{2}}.
\ee 
Substituting the ansatz into (\ref{krfe}), we find that the
only independent equations are the ones deriving from 
the $X^{m} = \{x, \theta_1, \theta_2, \theta_3 \}$ equations.
The equation deriving from $u$ follows from the $x$-equation,
and the remaining equations are satisfied trivially. 
This is expected as worldvolume diffeomorphisms can be used
to eliminate $p+1$ equations. The gauge field equation 
is satisfied automatically by the ansatz.
The independent equations are
\bea
x &:& \hspace{5mm} \del_{u} 
\left ( \frac{L_{\th_{\a}}} {(1 + u^4 (x')^2)^{\frac{1}{2}}} 
u^6 x' - q u^4 \right ) = 0; \label{set} \\
\theta_{\a} &:& \hspace{5mm}  
L_{\th_{\a}}^{-1} (1+ u^4 (x')^2)^{\frac{1}{2}}
\prod_{\b \neq \a} \sin^4 \theta_{\b} \sin^3 \theta_{\a} \cos \theta_{\a} 
= 0. \nonumber
\eea

\bigskip

To solve (\ref{set}) we first note that the angular equations
can be solved either when (i) all $\theta_{\a} = \half \pi$, 
which we will refer to as a maximal sphere, or when (ii)
one $\theta_{\a} = 0$ with the other two angles arbitrary,
which we will refer to as a minimal sphere.
The $x$ equation in (\ref{set}) yields
\be
x' = \frac{(qu^4 -c)}{\sqrt{ u^8 L_{\th_{\a}}^2 
- (qu^4 -c)^2}}, \label{xpm}
\ee
where $c$ is an integration constant. One can solve this 
differential equation using the first Appell hypergeometric 
functions of two variables, but we will not need this result.
Note that there is an unbroken translational invariance in
the $x$ direction. 

\subsection{Branes wrapping maximal spheres}

Let us now substitute solutions of the angular equations
into (\ref{xpm}). We focus first on the case of the brane wrapping a maximal
sphere. In this case, (\ref{xpm}) reduces to
\be
x' = \frac{(qu^4 -c)}{u^2 (u^8 + 2 c q u^4 - c^2)^{\frac{1}{2}}}, 
\label{xppm}
\ee
and the induced metric on the brane is
\be
ds^2 = u^2 (dx \cdot dx)_3 + \frac{u^6 (1+q^2)}{(u^4 - u_{+}^4)
(u^4 + u_{-}^4)} du^2 + (d\theta_4^2 + \sin^2 \theta_4 d\theta_5^2),
\ee
where we write
\be
(u^8 + 2cq u^4 - c^2) = (u^4 - u_{+}^4)(u^4 + u_{-}^4), \label{poly}
\ee
with $u_{+}^4, u_{-}^4 \ge 0$. The explicit form for the roots
of (\ref{poly}) are
\bea 
u^4_{+} &=& - cq + \left | c \right | \sqrt{1+q^2}; \\
u^4_{-} &=& cq + \left | c \right | \sqrt{1+q^2}. \nonumber
\eea

\subsubsection{$AdS_4 \times S^2$ embeddings}

The embedded geometry is $AdS_4 \times S^2$ when $c=0$.
In this limit (\ref{xppm}) integrates to the simple
expression
\be
x = x_{0} - \frac{q}{u}.
\ee
These embeddings were found by Karch and Randall \cite{KR}. Note
that the $AdS_4 \times S^2$ embeddings exist even in the zero 
flux ($q=0$) limit. The zero flux embedding must satisfy the 
zero extrinsic curvature trace condition (\ref{kcon}) since 
for this solution $J_{m}$ vanishes. It is a useful consistency check
on our equations and solutions to calculate explicitly the
extrinsic curvature for this embedding. The ${\cal K}^{k}_{ij}$
components vanish automatically since the extrinsic curvature
can be projected onto the normal bundle. 
The ${\cal K}^{x}_{ij}$ components
also vanish whilst for an $S^2$ embedded into $S^5$ we have
\be
{\cal K}^{\theta_{\a}}_{44} = 
{\cal K}^{\theta_{\a}}_{55} \sin^{-2} \theta_4 =
- \left [ \sin^2 \theta_{2} \sin^2 \theta_{3} 
\sin \theta_{1} \cos \theta_1, \sin^2 \theta_{3} \sin \theta_2 \cos \theta_2,
 \sin \theta_3 \cos \theta_3 \right ],
\ee
which implies that for this embedding the second fundamental
form vanishes (the embedding is totally geodesic), 
a stronger condition than (\ref{kcon}). This embedding has 
a very simple description as an intersection of a hyperplane
with the hyperboloid representing $AdS_5$ in a flat ambient
$6d$ spacetime with signature $(+,+,-,-,-,-)$.
We will return to this topic in section \ref{global}.

%\bigskip
\subsubsection{Asymptotically $AdS_4 \times S^2$ embeddings}

For the general solution in which $c \neq 0$ the induced
geometry is asymptotically $AdS_4 \times S^2$ for $u \gg u_{+}$.
Since from (\ref{xppm}) $x(u)$ becomes imaginary for $u < u_{+}$
the brane ends at $u = u_{+}$ ($x$ imaginary is not part of
the target space). The induced metric does not have singular
curvature at $u = u_{+}$: if we introduce a new coordinate 
$u = u_{+} + \rho^2$ with $\rho \ll 1$, then the metric in
this neighborhood becomes
\be
ds^2 = u_{+}^2 (dx \cdot dx)_{3} + \frac{u_{+}^3 (1+q^2)}
{(u_{+}^4 + u_{-}^4)} d\rho^2 + (d\theta_{4}^2 
+ \sin^2 \theta_{4} d\theta_{5}^2),
\ee
which is non-singular at $\rho = 0$. Geodesics
in the embedding geometry remain in the submanifold but 
they have finite endpoints at $u = u_{+}$; 
the embedded hypersurface is thus inextendible but incomplete.

One can bring the metric into a more conventional form
by changing variables from $u$ to the affine parameter $U$ of 
the radial geodesic:
\be
u^2 = \sqrt{U^{-4} - cq + \qu U^4 c^2 (1 + q^2)}.
\ee
This brings the metric into the form\footnote{
The metric (\ref{dwm}) is mapped to itself under the inversion
\be
U^4 \rightarrow {4 \over c^2 (1+q^2)  U^4}
\ee
and the range of $U$ becomes $0<U<U_+$.
Thus one can complete the spacetime by extending the 
$U$ variable to range from zero to infinity. The resulting manifold 
covers twice the original allowed region. This completion
corresponds to reflective boundary conditions for the geodesics 
that originally terminated at $u=u_+$. } 
\be
ds^2 = (1+ q^2) \frac{dU^2}{U^2} + 
\sqrt{U^{-4} - cq + \qu U^4 c^2 (1 + q^2)} (dx \cdot dx)_{3} 
+ (d\theta_{4}^2 + \sin^2 \theta_{4} d\theta_{5}^2). \label{dwm}
\ee
The range of $U$ is 
\be
U_{+} = \left ( \frac{2}{\left | c \right | \sqrt{1+q^2}} \right 
)^{\frac{1}{4}} \le U < \infty.
\ee
In AdS/CFT the radial coordinate corresponds to the energy scale,
which suggests that the dual theory develops a mass gap in the 
infrared. We will discuss the holographic interpretation of this 
embedding in section \ref{holdual}.

\subsection{Branes wrapping minimal spheres: D5-branes collapsing to D3 branes}

Let us now discuss embeddings in which the brane wraps
a minimal sphere. In this case (\ref{xpm}) reduces to
\be
x' = \frac{(qu^4-c)}{u^2 (2cq u^4 - c^2)^{\frac{1}{2}}}.
\ee
It is useful to rescale the parameter $c$ such that $c = C q$;
this removes all $q$ dependence in $x'$:
\be
x' = \frac{(u^4-C)}{u^2 (2C u^4 - C^2)^{\frac{1}{2}}}. \label{dif}
\ee
and the induced metric on the brane is then
\be
ds^2 = u^2 ( dx \cdot dx)_3 + \frac{u^6 du^2}{2 C (u^4 - \half C)}.
\ee
The embedding geometry is not asymptotically $AdS_4$: explicitly
integrating (\ref{dif}) for $u >>1$ we find that
\be
x \sim x_{0} + \frac{1}{\sqrt{2C}} u.
\ee
This implies that the defect in the dual field theory is
located at $x \rightarrow \infty$.
Since $x'$ becomes imaginary for $u^4 < \half C = u_c^4$, the brane ends
at $u_{c}$. The induced geometry is non-singular at $u_{c}$
though the embedded hypersurface is again incomplete. 

Note that the induced metric on the $S^2$ is degenerate. However,
provided that the flux through the sphere is non-zero
$M_{ij}$ is non-degenerate and hence invertible. Since the field equations
were derived in section 1 assuming the invertibility of $M_{ij}$,
but without assuming that $g_{ij}$ is non-degenerate, these
embeddings are admissible solutions of the field equations provided
that $q$ is non-zero. Examining the worldvolume action one finds
that the parameter $q$ appears only as an overall parameter and 
can thus be scaled to plus or minus one. 

Physically there is a very natural interpretation of 
embeddings in which the $S^2$ is minimal.
The D5-brane has effectively collapsed to a 
D3-brane embedded in $AdS_5$ and in fact this degenerate D5-brane
embedding can be found as a solution of the D3-brane equations
of motion. If the flux is positive we get a D3-brane, 
whereas negative flux corresponds to anti-D3-branes.
To show this, let us look for D3-brane
embeddings which preserve a $(2+1)$-dimensional
Poincar\'{e} invariance and which lie at a point in
the $S^5$. 

The D3-brane field equations in the $AdS_5 \times S^5$ target
space are
\be
\frac{1}{4!} \ep^{i_1i_2i_3i_4} f_{i_1 i_2 i_3 i_4 m} =
- \del_{i} (\sqrt{-M} G^{ij} \del_{j} X^{n} g_{mn})
+ \half \sqrt{-M} (G^{ij} \del_{i} X^{n} \del_{j} 
X^{p} g_{np,m}).
\ee
An appropriate ansatz for worldvolume coordinates is
\be
\xi^{i} = \{ x^0, x^1, x^2, u \},
\ee
whilst the transverse scalars are 
\be
X^{\lambda} = \{ x(u),\theta_{a} \}.
\ee
Then the only equation of motion (coming from the $u$ and $x$ field
equations, which are equivalent) is
\be
\del_{u} \left ( \frac{u^6 x'}{\sqrt{1+ u^4 (x')^2}} - u^4 
\right ) = 0. \label{d3fe}
\ee
The (constant) angles on the $S^5$ are arbitrary. 
Since the general solution of (\ref{d3fe}) is (\ref{dif}), 
this implies that the collapsed D5-brane wrapping
a minimal 2-sphere can be interpreted as a D3-brane. 

There are no special limits of (\ref{dif}) when $C=0$ or $q=0$. We cannot
solve the equations of motion for $C=0$ even when $q \neq 0$.
In the $q \to 0$  limit, $M_{ij}$ is totally
degenerate along the $S^2$ directions and the solution is
not admissible. 

We will see in section \ref{susymin} that the embedding 
discussed above is not supersymmetric. The reason is that the 
collapsed D5-brane
is a D3-brane which is misaligned with respect to D3-branes that 
create the background. Such branes would be expected to have 
an instability that tends to rotate them to
become aligned with the D3-branes creating the $AdS$ background.
The ansatz (\ref{ans1}) and (\ref{ans2}) used so far is not 
appropriate for finding such configurations. 
The appropriate ansatz describing D5-branes wrapping the $S^2$
and whose worldvolumes lie along $x$ is
\bea \label{xemb}
\xi^{i} &=& \{ x^0, x^1, x^2, x, \theta_{4}, \theta_{5} \}; \\
X^{\lambda} &=& \{ u, \th_1, \th_2, \th_3 \}; \nonumber \\
F_{\th_4 \th_5} &=& q \sin \th_4, \nonumber
\eea
where all transverse scalars are constant. The only field equations
which are not already satisfied by the ansatz are 
\bea
u &:& \hspace{10mm} u^3 (L_{\th_{\a}} - q) = 0; \\
\th_{\a} &:& u^4 L_{\th_{\a}}^{-1} 
\prod_{\beta \neq \a} \sin^4 \th_{\b} \sin^3 \th_{\a} 
\cos \th_{\a} = 0.  \nonumber
\eea
The only solution for which $M_{ij}$ is non-degenerate is an $S^2$ 
minimal solution with non-zero flux $q$ for any $u_{0}$.
As before, one can scale $q=\pm 1$.
We will see later that the solution with $q=1$
preserves 1/2 supersymmetry 
and can be interpreted as a supersymmetric D3-brane
whereas $q=-1$ breaks all supersymmetries and 
corresponds to an anti-D3-brane.

\section{Supersymmetry of embeddings} \label{kapsusy}
\noindent

Associated with every brane embedding is a kappa symmetry projection
which is defined (using quantities appearing in (\ref{dbiact})) 
as \cite{Cederwall:1996pv,Aganagic:1996pe,Cederwall:1996ri,
Bergshoeff:1996tu, Aganagic:1996nn, Bergshoeff:1997kr}
\be
d^{p+1} \xi \Gamma = - e^{-\Phi} {\cal{L}}_{DBI}^{-1} 
e^{\cal F} \wedge X |_{vol}, \label{kap1} 
\ee
with 
\be
X = \bigoplus_{n} \G_{(2n)} K^n I,
\ee
where $|_{vol}$ indicates that one should pick the terms proportional 
to the volume form, and the operations $I$ and $K$ act on spinors
 as $I \psi = -i \psi$
and $K \psi = \psi^{\ast}$. ${\cal{L}}_{DBI}^{-1}$  is the value of the
DBI Lagrangian evaluated on the background. Here we have used the notation 
\be
\G_{(n)} = \frac{1}{n!} d\xi^{i_{n}} \wedge ... \wedge d\xi^{i_1} 
\G_{i_1...i_n} ,
\ee
where $\G_{i_1...i_n}$ is the pullback for the target space
gamma matrices 
\be
\G_{i_1...i_n} =\del_{i_1} X^{m_1} ... \del_{i_n} X^{m_n} \G_{m_1...m_n}     
\ee
It has been shown in \cite{Cederwall:1996pv, Aganagic:1996pe, 
Cederwall:1996ri,Bergshoeff:1996tu, Aganagic:1996nn, Bergshoeff:1997kr} 
that $\G$ squares to one and 
is traceless. It follows that one can use $\G$ to project out 
half of the worldvolume fermions, thus equating the worldvolume 
fermionic and bosonic degrees of freedom. 
 
A given brane embedding within a supersymmetric target background preserves
some fraction of the supersymmetry provided that the Killing spinors
of the background $\ep$ are consistent with the projection
\be
\Gamma \ep = \ep. \label{projc}
\ee
In other words the restriction of the Killing spinors on the worldvolume 
should satisfy (\ref{projc}). We note here that (given our choice
of conventions) we will need to choose the positive sign in (\ref{projc})
in the AdS embedding, i.e. there are no supersymmetric 
D5-embeddings on $AdS_5 \times S^5$ with the negative sign.
 
To proceed we need the explicit form of the 
Killing spinors of the background.
The $AdS_5 \times S^5$ background geometry preserves maximal
supersymmetry since the dilatino equation is trivially satisfied and the
gravitino equation 
\be
\left ( D_{m} + \half i \g^{01234} \G_{m} \right ) \ep = 0, \la{eq1}
\ee
admits a full compliment of thirty-two independent solutions. 
(Conventions for the supersymmetry variations are given in the appendix.) 
We denote by $\g_a=e_a^m \G_m$ the tangent space gamma matrices.
For the case at hand, they are given by
\be
\g_p = {1 \over u} \G_p, \ (p=0,1,2,3), \quad \g_4 = u \G_u, \quad
\g_a= \left(\prod_{j=1}^{a-5} {1 \over \sin \theta_j} \right) 
\G_{\theta_{a-4}} \ (a=5,6,7,8,9) 
\ee
(when $a=5$ the product is equal to one).

Following closely Claus and Kallosh \cite{Claus:1998}, we now 
solve for the explicit
form of the Killing spinors. It is convenient to introduce
the projections:
\be
\ep_{\pm} = {\cal P}_{\pm} \ep = \half (1 \mp \g^{0123} I) \ep
=\half (1 \pm i \g^{0123}) \ep. 
\label{zproj}
\ee
Using these projectors we can rewrite the Killing spinor equations
(\ref{eq1}) as 
\bea
\del_{p} \ep_{-} + u \g_{p} \g_{4} \ep_{+} &=& 0; \nonumber \\
\del_{p} \ep_{+} &=& 0; \\
\del_{u} \ep_{\pm} \pm \half u^{-1} \ep_{\pm} &=& 0; \nonumber \\
D_{a} \ep_{\pm} \pm \half \g_4 \Gamma_{a} \ep_{\pm} &=& 0, \nonumber 
\eea
where $D_a$ is the covariant derivative on the sphere. The full
solution to the Killing spinor equation is the combination 
$\ep = \ep_{+} + \ep_{-}$ with
\bea
\ep_{+} &=& - u^{-\frac{1}{2}} \gamma_4 h(\theta_a) \eta_2; 
\nonumber \\
\ep_{-} &=& u^{\frac{1}{2}} h (\theta_a) (\eta_1 + x \cdot \gamma 
\eta_2),  \label{epp1}
\eea
where $\eta_1$ and $\eta_2$ are constant spinors, satisfying
\be
\eta_1 = {\cal P}_{-} \eta_1, \hspace{10mm} \eta_2 = {\cal P}_{+} 
\eta_2.
\ee
$\eta_{1}$ and $\eta_2$ are complex spinors of negative and
positive chirality respectively
so this gives us the $32$ independent real spinors. That is,
we can choose
\bea
\eta_{1} &=& \lambda - i \g^{0123} \lambda; \label{ep1} \\
\eta_2 &=& \eta + i \g^{0123} \eta, \nonumber 
\eea
where $\lambda$ and $\eta$ are real spinors of negative and 
positive chirality respectively,
 with $16$ independent components. Such a choice
makes the complex conjugation of the spinors manifest. The function
$h(\theta_a)$ appearing in both spinors results from the Killing
equation on the sphere and is given by
\be
h(\theta_a) = 
\exp (\half \theta_1 \g_{45}) \exp (\half \theta_2 \g_{56})
\exp (\half \theta_3 \g_{67})\exp (\half \theta_4 \g_{78})
 \exp (\half \theta_5 \g_{89}).
\ee
Explicit forms for the Killing spinors of $AdS_5 \times S^5$
appeared previously in, for example, \cite{Lu:1998}
and \cite{Claus:1998}. 

\subsection{Supersymmetry of asymptotically $AdS_4 \times S^2$ branes}

We would now like to check whether the asymptotically $AdS_4 \times 
S^2$ brane embedding found in the previous section preserves
supersymmetry. The explicit form of the kappa symmetry projection
is
\be
\ep = \frac{i}{u^4 (1+q^2)} \g^{012}  
\left ( (qu^4-c) \g^3 + (u^8 + 2 c q u^4 - c^2)^{\frac{1}{2}} \g^4 \right ) 
\left ( \g^{89} \ep^{\ast} - q \ep \right). \label{vcon}
\ee
For reasons discussed in Appendix B, we choose to work with
the projector $\Gamma$ involving the flux rather than to use
a similarity transformation to obtain a projector not involving
the flux $\Gamma'$ as in \cite{Bergshoeff:1997kr}; 
the difference compared to embeddings with flux considered previously is 
that here our worldvolume embedding depends explicitly on the flux. 

Preservation of supersymmetry requires that this condition must
be satisfied for some subset of the background Killing spinors
at all points on the brane worldvolume. In particular, it
must hold at all values of $x^p = (x^0,x^1,x^2)$. From the terms in
the Killing spinors which are linear in $x^p$ we find
the following condition
\bea
\g^{p} (1+ i \g^{0123}) h(\th_{a}) \eta &=& 
\frac{i \g^{012}}{u^4 (1 + q^2)} \left ( (qu^4 -c) \g^{3} + 
(u^8 + 2 c q u^4 - c^2)^{\frac{1}{2}} \g^4 \right ) \\
&& \times \left (\g^{89p}(1- i \g^{0123}) - q \g^{p}(1+i\g^{0123}) 
\right ) h(\th_{a}) \eta. \nonumber
\eea 
Recalling that $\eta$ is real, this can be separated into two conditions 
coming from the real and imaginary parts:
\bea
h(\th_{a}) \eta &=& \frac{1}{u^4 (1 + q^2)} \left ( (qu^4-c) - 
(u^8 + 2 cq u^4 - c^2)^{\frac{1}{2}} \g^{34} \right )(\g^{89} + q )
h(\th_{a}) \eta; \\
h(\th_{a}) \eta &=& - \frac{1}{u^4 (1 + q^2)} \left ( (qu^4-c) + 
(u^8 + 2 cq u^4 - c^2)^{\frac{1}{2}} \g^{34} \right )(\g^{89} - q )
h(\th_{a}) \eta, \nonumber
\eea
which in turn imply the constraints
\bea
\left (  (qu^4 -c) + q (u^8 + 2 c q u^4 - c^2)^{\frac{1}{2}} \g^{3489}
\right ) h(\th_a) \eta &=& 0; \label{sp1} \\
\frac{c}{qu^4} h(\th_{a}) \eta & =& 0. \nonumber
\eea
In this section we consider the case of non-zero $c$: then these
constraints can manifestly not be satisfied for non-zero
$\eta$. So setting $\eta = 0$ let us impose the kappa
symmetry projection on the remaining parts of the Killing
spinors involving $\lambda$. This implies the conditions
\bea
h(\th_{a}) \lambda &=& \frac{1}{u^4 (1 + q^2)} \left ( (qu^4-c) - 
(u^8 + 2 cq u^4 - c^2)^{\frac{1}{2}} \g^{34} \right )(\g^{89} + q )
h(\th_{a}) \lambda; \\
h(\th_{a}) \lambda &=& - \frac{1}{u^4 (1 + q^2)} \left ( (qu^4-c) + 
(u^8 + 2 cq u^4 - c^2)^{\frac{1}{2}} \g^{34} \right )(\g^{89} - q )
h(\th_{a}) \lambda, \nonumber
\eea
which impose constraints on $\lambda$ identical to those
in (\ref{sp1}). Thus $\lambda$ can also not be non-zero for
non-zero $c$ and the embeddings with $c \neq 0$ break all
the supersymmetry.

\subsection{Supersymmetry of $AdS_4 \times S^2$ branes}

When $c=0$ the kappa symmetry projection is given by
\be \label{kap}
\ep={i \over (1+q^2)} \g^{012} (q \g^3 + \g^4)(\g^{89} \ep^* - q \ep)
\ee
The restriction of the Killing spinors to the worldvolume is
\be
\ep=-u^{-1/2} (\g^4 + q \g^3) h(\theta_a) \eta_2 
+ u^{1/2} h(\theta_a) (x_0 \g^3 \eta_2 + \eta_1) 
+ u^{1/2} h(\theta_a) x^p \g_p \eta_2
\ee
The analysis of the previous section leading to (\ref{sp1})
still holds but since we now take $c=0$
the second condition in (\ref{sp1}) is trivially
satisfied and the first condition reduces to 
\be
(1+\g^{3489}) h(\theta_a) \eta = 0,
\label{j1}
\ee
Since $h(\theta_a)$ is invertible, (\ref{j1}) implies that half 
of the $\eta$ spinors are projected out.
To explicitly obtain the projection on the spinor $\eta$, 
we multiply (\ref{j1}) by $h(\theta_a)^{-1}$ 
and compute  $h(\theta_a)^{-1} \g^{3489} h(\theta_a)$.
This can be done effectively by using repeatedly identities of the form
\be
e^{-\half \theta \g_{p(p+1)}} \g_{(p+1)q} e^{\half \theta \g_{p(p+1)}}
=\cos \theta + \gamma_{qp} \sin \theta.
\ee
Recalling that $\th_1=\th_2=\th_3=\pi/2$ on the worldvolume,
we finally obtain
\be \label{etakil}
(1+\g^{3789}) \eta=0.
\ee
Let $\g^{3789} \eta_{\pm} = \pm \eta_{\pm}$. Equation (\ref{etakil})
eliminates the $\eta_+$ spinors.

We have just shown that the parts of the projection
condition (\ref{kap}) involving terms linear in $x^p$ can be satisfied
with $q$ arbitrary, provided we impose a projection
onto the constant spinor $\eta$. 
The projection condition (\ref{kap}) must be satisfied at
all points on the worldvolume, namely at all values of $u$. This
means that the projection holds independently for terms proportional
to $u^{\frac{1}{2}}$ and terms proportional to $u^{-\frac{1}{2}}$
in (\ref{kap}). The latter condition is automatically 
satisfied when (\ref{j1}) holds. Terms proportional to 
$u^{\frac{1}{2}}$ in (\ref{kap}) imply 
\be
(1 + \g^{3489}) h(\th_a) \lambda = - 2 \g^{3} x_{0} h(\th_a) \eta,
\ee
or equivalently
\be \label{lakil}
(1 + \g^{3789}) \lambda = - 2 \g^{3} x_{0} \eta.
\ee
Let $\g^{3789} \l_{\pm}=\pm  \l_{\pm}$. Equation (\ref{lakil}) 
determines $\l_+$ in terms of $\eta_-$, but leaves
undetermined $\l_-$.

Putting together all the projection conditions 
we see that in total sixteen of the Killing
spinors are preserved by the embedding and one half of
the target space supersymmetry is broken. 
Note that the projections on the constant spinors do not depend on
the flux. They do however depend on the asymptotic value
of $x = x_{0}$. Probe branes with different values
of $x_{0}$ (different locations of the defect in the boundary
theory) will preserve the same $\eta$ spinors but different
$\lambda$ spinors. Thus two or more defects in the boundary
theory will break the supersymmetry from one half to one
quarter. It would be interesting to understand this from
the perspective of the defect conformal field theory. 

\subsection{Supersymmetry of branes wrapping minimal spheres} \label{susymin}

The kappa symmetry projection for the brane wrapping a minimal
sphere is
\be
\ep = - i \g^{012} \left ( (1 - \frac{C}{u^4}) \g^{3} + 
\frac{1}{u^4} (2 C u^4 - C^2)^{\frac{1}{2}} \g^{4} \right ) \ep.
\ee
Preservation of supersymmetry requires that this condition must
be satisfied for some subset of the background Killing spinors
at all points on the brane worldvolume. From the terms in
the Killing spinors which are linear in $x^{p}$ we find that
\be
\g^{p} (1 + i \g^{0123}) h(\theta_a) \eta 
= - i \g^{012} \left ( (1 - \frac{C}{u^4}) \g^{3} + 
\frac{1}{u^4} (2 C u^4 - C^2)^{\frac{1}{2}} \g^{4} \right ) \g^{p}
(1 + i \g^{0123}) h(\th_a) \eta,
\ee
which can be separated into real and imaginary parts
\bea
h(\th_a) \eta &=& \left ( (1 - \frac{C}{u^4}) \g^{3} - 
\frac{1}{u^4} (2 C u^4 - C^2)^{\frac{1}{2}} \g^{34} \right )  
h(\th_a) \eta; \\
h(\th_a) \eta &=& \left ( (1 - \frac{C}{u^4}) \g^{3} + 
\frac{1}{u^4} (2 C u^4 - C^2)^{\frac{1}{2}} \g^{34} \right )  
h(\th_a) \eta; \nonumber,
\eea
which can only be satisfied if
\be
\frac{C}{u^4} h(\th_a) \eta = 0; \hsp 
\g^{34} (2 C u^4 - C^2)^{\frac{1}{2}} h(\th_a) \eta = 0, 
\label{conw}
\ee
conditions which do not have non-zero solutions $\eta$ when 
C is non-zero (and recall that $C$ cannot be zero). Thus we
must set $\eta = 0$. Imposing the kappa symmetry projection
on the remaining parts of the Killing spinors involving 
$\lambda$ we find constraints on $\lambda$ identical to those
in (\ref{conw}). Thus there are no non-zero solutions to
the kappa symmetry projection and these embeddings
break all the supersymmetry. 

Finally we discuss the D5-brane embedding 
described in (\ref{xemb}). The kappa symmetry projection reads
\be \label{d3pr}
\e = - i sgn(q) \g^{0123} \e
\ee
where $sgn(q)$ is the sign of flux (which as we  argued 
can be scaled to $\pm 1$). 
This equation should hold at all points on the worldvolume.
Inserting the Killing spinors from (\ref{epp1}) and examining 
the terms linear in $x^p$ we find 
that the resulting equation is satisfied identically when $sqn(q)=1$,
but it projects out the $\eta$ spinor when $sqn(q)=-1$.
The remaining equations project out the $\eta$ spinor when $sqn(q)=1$,
and the $\l$ spinor when $sqn(q)=-1$. We conclude that 
the embedding with $sqn(q)=1$ preserves half of supersymmetry
and can be identified with a supersymmetric D3-brane 
whilst the embedding with $sqn(q)=-1$
breaks all supersymmetry and corresponds to an anti-D3 brane.

\section{Dual interpretation: RG flows on the defect} \label{holdual}

D5-branes wrapping 
submanifolds of $AdS_5 \times S^5$ may be viewed
as the near-horizon limit of intersecting D3-D5 systems.
The AdS/CFT duality is considered to act twice, both in the bulk and
on the worldvolume of the D5-brane. The dual field theory could
be obtained directly by considering the intersections of
the D3-brane and D5-brane worldvolume theories: in the case
we discuss here it will be $d=4\ {\cal N}=4$ SYM theory coupled to 
a three dimensional defect. The defect theory may be associated with
the boundary of the $AdS_4$ of the $AdS_4 \times S^2$ D5-brane and as 
such it should be a conformal field theory. The defect theory
contains both ambient fields, which follow from the $d=4\ {\cal N} = 4$
SYM, and fields confined to the defect.
An explicit construction of this defect CFT theory was given
recently in \cite{ooguri1}, see also 
\cite{Sethi:1997zz,Ganor:1997jx,Kapustin:1998pb,Hori:2000ic} 
and \cite{johanna}.  
 
Within the approximations used in 
this paper, the D5-brane theory does not backreact on the 
bulk. This means that we can consider deformations in which 
the boundary theory remains conformal but the defect theory runs. 
This is precisely what is happening in 
our asymptotically $AdS_4 \times S^2$ embeddings into the
$AdS_5 \times S^5$ background. 

Since in our embeddings 
only the $AdS_4$ part of the solution is deformed
this implies that the defect QFT still has the same R-symmetry
as the defect CFT. Using the operator-field dictionary 
developed in \cite{ooguri1} and the asymptotic form of the 
worldvolume fields we can show that the RG-flow corresponds 
to a vev deformation of the defect theory. To see this, note
that the active scalar in our embeddings behaves for large $u$ as
\be
u x \sim u x_{0} + \frac{c}{5 u^4},
\ee
where we assume that $q=0$ (since this is the only case considered
in \cite{ooguri1}). Using the standard  
AdS/CFT dictionary \cite{maldacena}, \cite{gubser}, \cite{witten},
\cite{ooguri2}, this suggests that this scalar is dual
to an operator in the defect theory of conformal dimension four.

The identification between the scalar $ux$ and such an operator was 
made in \cite{ooguri1}, where it was shown that the operator
in question is a certain four supercharge descendant of the second
floor chiral primary on the defect. Let us briefly summarise
their arguments; for more details we refer to \cite{ooguri1}. 

The defect theory has an $SU(2)_{H} \times SU(2)_{V}$ R-symmetry;
the first factor is associated with rotations of the $S^2$ wrapped
by the D5-brane whilst the second is associated with the symmetry
of transverse directions. The ambient fields on the defect follow
from the decomposition of the ${\cn = 4}$ vector multiplet into a
3d ${\cal{N}} = 4$ vector multiplet and a 3d ${\cn} = 4$ adjoint
hypermultiplet. The bosonic components of the former are $(A_{k}, X^{A}_{V})$
and of the latter are $(A_3, X^{I}_{H})$, where we use $k=0,1,2$ to denote
spacetime indices, $A$ denotes a vector index of $SU(2)_{V}$ and $I$
denotes a vector index of $SU(2)_{H}$. 

There is also a $d=3$ hypermultiplet on the defect which transforms
in the fundamental of the $SU(N)$ gauge group. This consists of
an $SU(2)_{H}$ doublet of complex scalars $q^{m}$ and an $SU(2)_V$
doublet of Dirac fermions $\psi^{i}$. It was argued in \cite{ooguri1} 
that the lowest chiral primary of the defect theory is the triplet
\be
C^{I} \equiv \bar{q}^{m} \sigma^{I}_{mn} q^{n}.
\ee
T-duality arguments were then used to show that higher chiral
primaries must arise from taking the symmetrised traceless
part of this operator with the scalars from the ambient
hypermultiplet, namely
\be
C^{I_1..I_{l}} = C^{(I_1} X_{H}^{I_2}...X_{H}^{I_{l})}.
\ee
The $l=2$ chiral primary has $\Delta = 2$; thus it must 
have a four supercharge descendant ${\cal O}_{x}$ 
which has $\Delta = 4$ and is an R-singlet. 

Now the AdS/CFT dictionary tells us that 
we should regard $x_{0}$ as a source for
the operator ${\cal O}_x$, and $c$ as its expectation value. Since
the operator that gets the vev is a specific scalar component of 
a $3d$ superfield, the vev breaks supersymmetry, as we found.
Furthermore, the operator is an R-singlet in accordance 
with the fact that the $S^2$ part of the solution is undeformed.
Thus our results provide evidence for the dictionary proposed
in \cite{ooguri1}. 

\bigskip

An interesting feature of this RG-flow is that the theory
develops a mass gap in the infrared. We will now suggest a 
way to understand this, based on extending our embeddings to
embeddings of a probe D5-brane in the full D3-brane background. 
Let us write the metric in the D3-brane background as 
\be
ds^2 = f(r)^{-\frac{1}{2}} (dx \cdot dx)_4 + 
f(r)^{\frac{1}{2}} (dr^2 + r^2 d\Omega_5^2),
\ee
where $f(r) = (1 + R^4/r^4)$. Then the N D3-branes are located
at the origin in the transverse space. Standard intersection
rules tell us that a probe D5-brane intersecting the D3-branes
on a membrane will preserve supersymmetry. This corresponds
to taking the worldvolume directions of the D5-brane to 
be $\{ x^0,x^1,x^2,r,\Omega_2 \}$, where the brane wraps
a maximal $S^2$ in the $S^5$ and all other transverse scalars,
including $x^3 \equiv x$, are constant. Taking the near horizon limit,
$r \ll R$, reproduces the $AdS_4 \times S^2$ embeddings
considered here. The probe D5-brane intersects the
D3-branes on a submanifold on which $r=0$ and $x= x_{0}$; this means
that there are massless 3-5 strings. 

\begin{figure}[hbt]
\centerline{ \epsfxsize 3in \epsfbox{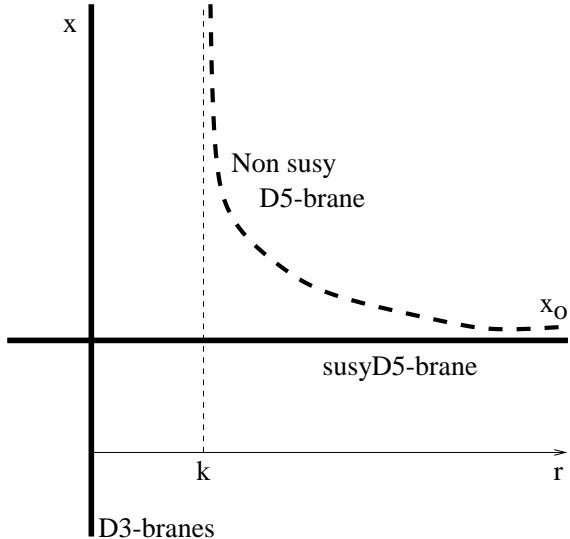} } \caption[]{
Probe D5-branes in the D3-brane background
} \label{figone}
\end{figure}

Suppose we now look for an embedding in which $x$ depends explicitly
on $r$; the analysis follows closely that of the section three
and leads to a defining equation for $x(r)$:
\be
0 = \frac{\partial}{\partial r} \left ( \frac{ x'(r) f^{-1}(r)}
{\sqrt{(1 + f(r)^{-1} (x'(r))^2)}} \right ), \label{sxc}
\ee
which in the near horizon limit reproduces (\ref{xppm}). Using
the explicit form of the Killing spinors in the D3-brane background
one can show that such an embedding breaks supersymmetry unless
$x$ is constant (as expected from our near horizon analysis). 
Furthermore, examining the asymptotics of the general solution 
to (\ref{sxc}), we find
that the schematic dependence of $x(r)$ is as illustrated in 
Figure \ref{figone}. 

The key point is that $r \ge k$ (where $k \ge 0$ corresponds to the $c$
appearing in (\ref{xppm}))
and so the non-susy probe D5-brane does not intersect the D3-branes:
$k$ measures the separation of the probe from the D3-branes.
All 3-5 strings are massive and this is the origin of
the mass gap in the defect quantum field theory. 
This is consistent with the fact that the operator 
$\co_x$, which we argued gets a vev, contains fields 
of the defect hypermultiplet.

It would be interesting to further explore this holographic duality 
by computing correlation functions and Wilson 
loops \cite{Maldacena:1998im,Rey:1998ik}. To properly 
compute correlation functions one would need to implement the 
program of holographic renormalization 
\cite{deHaro:2000xn,Bianchi:2001de,Bianchi:2001kw}
in the current setting. We leave this problem for future work.

\section{Penrose limits} \label{penrose}

In this section we consider the Penrose limit of
$AdS_5 \times S^5$ leading to a pp-wave and the limit thus 
induced on the $AdS_4 \times S^2$ brane embeddings.

\subsection{Embeddings in global coordinates} \label{global}

As is well-known, $AdS_5$ can be described as a pseudosphere
embedded in a 6-dimensional ambient space. 
Introducing coordinates $Y^{\mu}$ for this ambient space, then
\be
(Y^0)^2 + (Y^1)^2 - (Y^2)^2 - (Y^3)^2 - (Y^4)^2 - (Y^5)^2 = R^2, 
\label{cart}
\ee
where $R$ is the curvature of the $AdS_5$ hypersurface. 
Global coordinates for $AdS_5$ are
related to the Cartesian coordinates $Y^{\mu}$ as
\bea
Y^{0} = R \cosh \rho \cos \tau; \hsp & \hsp
Y^{1} = R \cosh \rho \sin \tau; \nonumber \\
Y^{2} = R \sinh \rho \cos \chi \sin \psi; \hsp & \hsp
Y^{3} = R \sinh \rho \cos \chi \cos \psi; \label{cart1}\\
Y^{4} = R \sinh \rho \sin \chi \sin \phi; \hsp & \hsp
Y^{5} = R \sinh \rho \sin \chi \cos \phi, \nonumber
\eea
and the $AdS_5$ metric in these coordinates is
\be
ds^2_5 = R^2 \left [ - \cosh^2 \rho d\tau^2 + d \rho^2 
+ \sinh^2 \rho (d\chi^2 + \cos^2 \chi d\psi^2 
+ \sin^2 \chi d\phi^2) \right ]. \label{glob}
\ee
We could now repeat the analysis of the previous sections
to find $AdS_4 \times S^2$ embeddings in this coordinate 
system, by solving the D-brane equations of motion explicitly.
It is convenient however to use a different approach. 
We found in the Poincar\'{e} coordinate system that the 
supersymmetric branes wrapped an $AdS_4$ submanifold, of 
curvature radius $R \sqrt{1+q^2}$ where $q$ is the charge on the $S^2$. 
This should be a coordinate independent statement. 

\bigskip

To find such $AdS_4$ submanifolds in global coordinates
it is most convenient to
start from the Cartesian embedding coordinates \footnote{This
approach was also used in \cite{bachas2} to find $AdS_2$ submanifolds
of $AdS_3$ in different coordinate systems.}. Suppose
we choose the codimension one hypersurface $Y^2 = R q$ in the
ambient space; then inserting this condition into 
(\ref{cart}) we find the intersection of this hypersurface
with the $AdS_5$ hypersurface is a $4$-dimensional hypersurface
satisfying
\be
(Y^0)^2 + (Y^1)^2 - (Y^3)^2 - (Y^4)^2 - (Y^5)^2 = R^2 (1 + q^2).
\ee
This implies that the $4$-dimensional hypersurface is also
$AdS$, with curvature radius $R \sqrt{1+ q^2}$. Other $AdS_4$
submanifolds can be obtained by choosing different codimension
one hypersurfaces; the submanifolds are related to each other
by the action of the five-dimensional isometry group $SO(4,2)$. 

The induced metric on the hypersurface $Y^2 = Rq$ in $AdS_5$
can be written in terms of the global coordinates as
\bea
ds^2_4 &=& R^2 \left ( - \cosh^2 \rho d\tau^2 + d \rho^2 
+ \sinh^2 \rho (d\chi^2 + \sin^2 \chi d\phi^2) \right) 
\label{ads4} \\
&& + \frac{R^2 q^2}{(\sinh^2 \rho \cos^2 \chi - q^2)}
(\sinh \rho \sin \chi d\chi - \cosh \rho \cos \chi d\rho)^2, 
\nonumber
\eea
where we eliminate $\psi$ in favour of $(\rho,\chi)$. 

This submanifold defines a brane embedding in which the worldvolume
coordinates of the $AdS_4$ 
are $(\tau,\rho,\chi,\phi)$ with $\psi$ a transverse scalar given by
\be
\psi = \arcsin (\frac{q}{\sinh \rho \cos \chi}).
\ee 
Note that $\psi = 0$ for zero flux embeddings ($q=0$) and
that $\psi \rightarrow 0$ as $\rho \rightarrow \infty$ for
finite flux embeddings. 

Having found an $AdS_4$ embedding in global coordinates 
by this quick route,
one can then verify that it does indeed satisfy the D5-brane equations
of motion, with the D5-brane wrapping a maximal $S^2$ with
flux $q$. Furthermore, one can check explicitly that the curvature 
scalar of (\ref{ads4}) is indeed
\be
{\cal R} = \frac{12}{R^2(1+q^2)}.
\ee
One could also use an explicit relationship between
Poincar\'{e} and global coordinates to map
this embedding to that found previously. 

\subsection{Penrose limits}

The reason for switching to global coordinates is that there is then
a  particularly simple Penrose limit taking $AdS_5 \times S^5$ to 
an Hpp-wave. The background is
\bea
ds_{10}^2 &=&  ds^2(AdS_5) 
+ R^2 \left(d\theta_1^2 + \sum_{k=2}^5 \prod_{j=1}^{k-1}  
\sin\theta_j^2 d\theta_k^2\right); \\
f_{5} &=& 4 R^4 \sinh^3 \rho \cosh \rho d\tau \wedge d\rho \wedge
d\Omega_3
+ 4 R^4 d\Omega_5, \nonumber
\eea
where the $AdS_5$ metric is given in (\ref{glob}). 
A Penrose limit focuses on the geometry in the neighbourhood
of a null geodesic \cite{penrose}, \cite{guven}. 
We consider here limits
arising from a particle moving in the $\th_{5}$ direction,
sitting at $\th_{a} = \half \pi$ for $a \neq 5$. Let us introduce
coordinates
\be
\th_{a} = \half \pi - \frac{ z^{a}}{R},
\ee
for $a \neq 5$; the Penrose limit requires $R \rightarrow \infty$ so
expanding the metric on the $S^5$ for large $R$ we find
\be
ds_{5}^2 = (dz^a)^2 + (R^2 - (z^a)^2) d\th_5^2 +...,
\ee
where the ellipses denotes terms in negative powers of $R$. 
Introducing coordinates
\be
x^{+} = \half (\tau + \th_{5}); \hsp  x^{-} = \half  
R^2 (\tau - \th_{5}); \hsp \rho = \frac{r}{R}, \label{pen}
\ee
and then taking the limit $R \rightarrow \infty$ the target
space metric becomes \cite{blau0}, \cite{BMN}
\be
ds^2 = - 4 (dx^{+} dx^{-}) - ((y^a)^2 + (z^a)^2) (dx^+)^2 + (dz^a)^2 
+ (dy^a)^2 
\ee
which is a plane wave metric, whilst the flux becomes
\be
f_{+ y^1 y^2 y^3 y^4} = f_{+ z^1 z^2 z^3 z^4} = 4,
\ee
where we introduce for convenience below Cartesian coordinates
such that $y^a y^a = r^2$,
\be \label{ycoor}
y^1=r \sin \chi \cos \f, \quad y^2 = r \sin \chi \sin \f, \quad 
y^3=r \cos \chi \cos \psi, \quad y^4 = r \cos \chi \sin \psi. 
\ee

Now let us consider the Penrose limit applied to the worldvolume
fields. The induced worldvolume metric is
\be
ds^2_6 = ds_4^2 + R^2 (d\th_4^2 + \sin^2 \theta_4 d\th_5^2),
\ee
where the $AdS_4$ metric is given in (\ref{ads4}). There is
also a flux on the $S^2$, $F_{\th_4 \th_5} = q \sin \th_4$.  
The Penrose scaling of the induced worldvolume metric
is implicitly defined given the scaling of the target space
metric \cite{blau0}, \cite{blau}. 

The scaling of the worldvolume flux is determined
by requiring that any solution of the D-brane field equations
remains a solution in the Penrose limit. As discussed in
\cite{blau} this means that the flux on the brane in
the Penrose limit $\tilde{F}$ is related to the original
flux $F$ by $\tilde{F} = R^2 F$. Since in the Penrose limit
\be
F \rightarrow \frac{q}{R} dz^{4} \wedge d\th_5,
\ee
a finite value for $\tilde{F}$ requires that we set 
$\tilde{q} = q R$.  

Then applying this Penrose limit to the induced
worldvolume metric in (\ref{ads4}) we get
\bea
ds^2_6 &=&  - 4 dx^{+} dx^{-} 
- ((y^1)^2 + (y^2)^2 + (y^3)^2 + \tilde{q}^2 + (z^4)^2) (dx^+)^2 \\
&& + (dz^4)^2 + (dy^1)^2 + (dy^2)^2 + (dy^3)^2, \nonumber
\eea
where the $y^i$ coordinates are defined in (\ref{ycoor}).
This is a pp-wave brane located at $y^4 = \tilde{q}$
and $z^1 = z^2 = z^3 = 0$. The rescaled worldvolume flux
is
\be
\tilde{F} = \tilde{q} dz^4 \wedge dx^{+}.
\ee
Different initial choices for
the $AdS_{4}$ submanifold will lead to branes located at
codimension one hypersurfaces in the $y^a$ plane.
Since all such branes wrap the same
maximal $S^2$ in the $S^5$, their Penrose limits will always
lead to branes located at $z^1 = z^2 = z^3 =0$.

\subsection{Branes in the plane wave background} \label{ppbr}

Following the general theme of this paper, it is interesting to find
the plane wave brane embeddings into the Hpp-wave background
directly from the D-brane equations of motion and to derive their
preserved supersymmetries explicitly. 
So let us look for D5-brane embeddings in which the worldvolume
coordinates are 
\be
\xi^{i} = \left \{ x^{+}, x^{-}, y^1, y^2, y^3, z^4 \right \},
\ee
and the transverse scalars are
\be
X^{\lambda} = \left \{ y^{4}, z^{1}, z^{2}, z^{3} \right \},
\ee
and do not depend on the worldvolume coordinates. We also
assume that there is a flux $\tilde{F} = \tilde{q} dz^4 \wedge dx^{+}$. 
Then the D-brane equations of
motion from (\ref{finfe}) reduce to 
\bea
\del_{i} (\sqrt{-M} \th^{ij}) = 0; \nonumber \\
0 = - \del_{i} (\sqrt{-M} G^{ij} \del_{j} X^{n} g_{mn}) 
+ \half (\sqrt{-M} G^{ij} \del_{i} X^{n} \del_{j} X^{p} g_{np,m}). 
\label{zfeq}
\eea
since there are no WZ source terms.
Substituting in the ansatz, we find that these equations impose
no further constraints: to check this note that $\sqrt{-M} = 2$,
$G^{++} = 0$ and the only non-zero component of $\th^{ij}$ is
$\th^{-z^4} = \tilde{q}/2$. The constant values of the transverse
scalars are left arbitrary (effectively because $G^{++} = 0$). 

However, the Penrose limit of the supersymmetric $AdS_4 \times 
S^2$ branes gives D5-branes located at $(y^4)^2 = \tilde{q}^2$
and $z^1 = z^2 = z^3 = 0$. 
A natural question to ask is thus whether the plane wave
brane is supersymmetric for arbitrary values of the transverse
scalars or just for these values. To check this we will use
the kappa symmetry projection.

\bigskip

Let us first construct the target space Killing spinors, following
closely the analysis given in \cite{blau0}. 
Choose the vielbein to be
\be
e^{\hat{-}} = 2 dx^{-} + \half ((y^a)^2 + (z^a)^2) dx^{+}; \hsp
e^{\hat{+}} = - dx^{+}; \hsp  
e^{\hat{a}} = dy^{a}; \hsp
e^{\hat{(a+4)}} = dz^a, 
\ee
where we denote tangent space indices by hats. This implies
the choice for the tangent space metric component 
$\eta_{\hat{-} \hat{+}} = 1$. The target space
Dirac matrices $\G$ can then be expressed as
\bea
\G^{-} = \half \g_{+} + \qu ( (y^a)^2 + (z^a)^2 ) \g_{-}; \hsp
\G^{+} = - \g_{-}; \hsp 
\G^{y^a} = \g_{a}; \hsp
\G^{z^a} = \g_{(a+4)}.
\eea
Using the gravitino supersymmetry transformations we find that
the target space Killing spinors satisfy $\del_{-} \ep =0$ and
\bea
(\del_{y^a} + \half i \g_{- {a} 1234}) \ep = 0; \hsp
(\del_{z^a} + \half i \g_{- (a+4) 5678}) \ep &=& 0; \\
(\del_{+} + \half \sum_{a=1}^{4} 
(\g_{- a} y^{a} + \g_{-(a+4)} z^a) +  \half i (\g_{1234} 
+ \g_{5678})) \ep &=& 0. \nonumber
\eea
To derive these equations we have used the fact that
$\ep$ is a negative chirality spinor, so that
\be
\g_{+-12345678} \ep = - \ep.
\ee
The solution to these equations is
\bea
\ep &=& (1 - \half i \sum_{a=1}^{4} \g_{-} (y^a \g_{a} \g_{1234} + 
z^a \g_{(a+4)} \g_{5678})) 
(\cos(\half x^{+}) - i \sin( \half x^{+}) \g_{1234}) \label{ppspin} \\
&& \hspace{10mm} \times (\cos(\half x^{+}) 
- i \sin( \half x^{+}) \g_{5678}) (\lambda + i \eta), \nonumber
\eea
where $\lambda$ and $\eta$ are constant real negative chirality spinors. 

\bigskip

The kappa symmetry projection is
\be
\ep = i \left ( \g_{+-1238} \ep^{\ast}-\tilde{q} \g_{-123} \ep \right)
\ee
to be evaluated on the embedding hypersurface which has
constant $(y^4,z^1,z^2,z^3)$. The condition must hold
at all values of $x^{+}$ (since this is 
a worldvolume coordinate) and so using the explicit form
of the Killing spinors (\ref{ppspin}) leads to the conditions
\bea
(1 + iR) (1 + iP) (\lambda + i \eta) &=& i Q (1 - i P) (\lambda - i \eta);
\label{spc1} \\
(1 + i R) (1 + iP) (\g_{1234} + \g_{5678})
(\lambda + i \eta) &=& - i Q (1 - i P) (\g_{1234} + \g_{5678}) 
(\lambda - i \eta); \nonumber \\
(1 + iR)
(1 + iP) \g_{12345678} (\lambda + i \eta) &=& i Q (1 - i P) 
\g_{12345678} (\lambda - i \eta), \nonumber
\eea
where
\bea
Q &=& \g_{+ -1238}; \label{zsc1} \\
P &=& - \half \sum_{a=1}^{4} \g_{-} (y^{a} \g_{a} \g_{1234} + z^{a} 
\g_{(a+4)} \g_{5678}); \nonumber \\
R &=& \tilde{q} \g_{-123}. \nonumber
\eea
The real and imaginary parts of the first and third conditions 
in (\ref{spc1}) imply
\bea 
(QP-1) \lambda + (P+Q+R) \eta &=& 0; \\
(P+R -Q) \lambda + (1 + QP) \eta &=& 0; \nonumber \\
(1 + QP) \lambda + (P+R - Q) \eta  &=& 0; \nonumber \\
(P+Q+R) \lambda + (QP-1) \eta &=& 0, \nonumber 
\eea
where we have used $P^2 = 0 = RP$ (since $\g_{-}^2 = 0$). 
These equations then imply the conditions
\be 
\lambda = Q \eta; \hsp 
\{ P, Q \} \eta = - Q R \eta; \hsp [Q, R] \eta = 0. \label{cnc0}
\ee
The first of these conditions breaks the supersymmetry by one
half. The third condition is satisfied automatically since $Q$ 
and $R$ commute. 
Explicitly evaluating the anticommutator, the second condition
requires that 
\be
(\g_{-8} y^4 - \g_{-12367} z^1 + \g_{-123 57} z^2 - \g_{-12356} z^3)
\eta = - \tilde{q} \g_{-8} \eta. \label{cnc1}
\ee
This condition can be satisfied in two ways. One possibility is
$y^4 = - \tilde{q}$ with $z^1 = z^2 = z^3 = 0$, in which case no
condition needs to be imposed on $\eta$. We note here that 
had we chosen the negative sign in the kappa symmetry 
projection (cf equation (\ref{projc})), we would still 
have gotten a supersymmetric configuration, but located
at $y^4 = \tilde{q}$ and $z^1 = z^2 = z^3 = 0$. One 
should contrast this to the case of the $AdS$ embeddings
where only the positive sign in the kappa symmetry projection
yielded a supersymmetric solution.

We still need to check that the second condition in (\ref{spc1}) 
is satisfied. Using
the negative chirality of $\eta$ and $\lambda$ we can rewrite
the condition as
\be
(1 + i R) (1 + iP) \g_{1234} \g_{-} \g_{+} (\lambda + i \eta)
= - i Q (1 - i P) \g_{1234} \g_{-} \g_{+} (\lambda - i \eta).
\ee
Again using $R \g_{-} = P \g_{-} = 0$ this reduces to
\be
\g_{1234} \g_{-} \g_{+} (\lambda + i \eta)
= - i Q \g_{1234} \g_{-} \g_{+} (\lambda - i \eta) 
= i \g_{1234} \g_{-} \g_{+} Q (\lambda - i \eta),
\ee
which is manifestly satisfied when $\lambda = Q \eta$
as in (\ref{cnc0}). The brane
embedding then breaks the background supersymmetry by one half,
the projection on the spinors being
\be
\lambda = \g_{+ -1238} \eta. 
\ee
This is what happens for the Penrose limits of $AdS_4 \times S^2$ 
branes. 

The second possibility to satisfy (\ref{cnc1})
is that we impose the constraint on $\eta$
\be
\g_{-} \eta = 0.
\ee
The constraint can then be satisfied for arbitrary 
$(\tilde{q}, z^1, z^2, z^3, y^4)$. The second condition in (\ref{spc1})
is then automatically satisfied and the embeddings then break the
background supersymmetry to one quarter. 

\section{Other $AdS_{m+1} \times S^{n+1}$ branes and 
their Penrose limits} \label{general}

Before going on to discuss more generally branes in pp-wave backgrounds
we would like to consider other Dp-brane embeddings
in $AdS_5 \times S^5$ and their Penrose limits. 
As we pointed out in section two, provided that there
are no WZ source terms, the Dp-brane equations of motion
reduce to the constraint that the trace of
the second fundamental form of the embedding is zero. 
Many such embeddings will exist but we want to focus on
embeddings of the form $AdS_{m+1} \times S^{n+1}$, which
originate from intersections of D$(m+n+1)$-branes with the
background D3-branes. 
All such embeddings are totally geodesic provided the sphere is
maximal (the
second fundamental form vanishes) and hence they satisfy
the equations of motion. The most economic way to explicitly
verify this is to work in Poincar\'{e} coordinates and choose
an ansatz 
\bea
\xi^{i} &=& \{ x^0,..,x^{m}, u, \th_{5-n},.., \th_5 \}; \nonumber \\
X^{\lambda} &=& \{ x^{m+1},..,x^{3}, \th_1,...,\th_{4-n} \}.
\eea
The equations of motion are then satisfied provided that
the transverse scalars in the $AdS_5$ are constant and 
the wrapped sphere is maximal. Note that
the $AdS_2 \times S^4$ embedding was already found in 
\cite{Pawelczyk:2000hy}, and can be
generalised in an obvious way by putting electric flux on the $AdS_2$.

The dual interpretation of these branes has not been discussed
beyond the $AdS_4 \times S^2$ branes considered in detail here
but they should all be understood in terms of higher codimension
defects in the field theory. One could derive the effective field
theory from appropriate intersections of the (flat space)
D$(m+n+1)$-brane and D3-brane worldvolume theories. 
All these defects should preserve
a subgroup of the conformal invariance of the bulk field theory
because of the conformal invariance of the induced worldvolume metrics. 
We discuss the dual dCFTs further in \S\ref{gauD}.

\bigskip

Consider next the supersymmetry of these embeddings. The
kappa symmetry projectors are
\bea
\G_{(m+1),(n+1)} = \g^{0..m 4 (9-n).. 9} K^{\frac{m+n+2}{2}} I
\eea
where we recall that $K$ acts by complex conjugation, 
$I$ by a multiplication by $-i$. (We could
of course just write $K = \ast$ but sticking to this notation makes
the relation with other spinor conventions more manifest.) 
One follows similar analysis to that given for $AdS_4 \times 
S^2$ branes to demonstrate that the these branes are one half
supersymmetric for $p = 1,5$ when both $m$ and $n$ are odd, whilst 
they are supersymmetric for $p = 3,7$ when both $m$ and $n$ are even.
This gives rise to the possibilities listed in Table 1, namely
$AdS_2$, $AdS_3 \times S^1$, $AdS_4 \times S^2$, 
$AdS_2 \times S^4$, $AdS_5 \times S^3$ and $AdS_3 \times S^5$.
The key point of the analysis is that preservation of supersymmetry
requires that
\be
[ {\cal P}_{\pm}, \Gamma_{(m+1)(n+1)} ] = 0,
\ee
where ${\cal P}_{\pm}$ are the projections introduced in (\ref{zproj}). 
One can easily show that this condition is satisfied by
$\G_{(m+1),(n+1)}$ only in the cases we list above. As we discuss
in the introduction, the supersymmetric $AdS_{m+1} \times S^{n+1}$
embeddings are in one to one correspondence with the near horizon
limits of supersymmetric intersections of D3-branes with other 
Dp-branes, as one would expect. 

\bigskip

Now let us take the Penrose limits of these brane embeddings. We
could do this explicitly by writing the embeddings in
terms of global coordinates for the background and then applying
the appropriate Penrose limit. However, from the $AdS_4 \times S^2$ case
we can already see the pattern. Provided that the brane wraps
the boosted circle, the $AdS_{m+1} \times S^{n+1}$
brane will be mapped to a pp-wave D(m+n+1)-brane with induced metric
\be
ds^2 = - 4 dx^{+} dx^{-} - \left ( \sum_{a=1}^{m}(y^a)^2 + 
\sum_{a=1}^{n}(z^a)^2 \right) (dx^+)^2 
 + \sum_{a=1}^{m} (dy^a)^2 + \sum_{a=1}^{n} (dz^a)^2, 
\ee
where the transverse positions are all zero. As we will see below  
branes along the light
cone always preserve at least one quarter of the supersymmetry, even
when located at arbitrary transverse positions. They
preserve one half of the supersymmetry only in the special cases
corresponding to the one half supersymmetric branes in $AdS_5 \times
S^5$ listed above, namely when the branes are located at the origin
and either $p = 1,5$ with $(m,n)$ both odd or $p=3,7$ with $(m,n)$ 
both even. The same results were  obtained recently in \cite{dabholkar}
by analyzing open strings in the pp-wave background.
Note that the $AdS_2$ brane cannot wrap the boosted circle
and cannot be mapped to a lightcone brane. We will discuss later
the Penrose limits of branes which are orthogonal to the boosted circle.

\section{Branes in the pp-wave background} \label{branes}

Now let us discuss more generally Dp-brane embeddings in the pp-wave
background and their supersymmetry. In this section we find
all possible brane embeddings in which the transverse scalars are
constants and there is zero worldvolume flux. It is convenient
to discuss separately branes with two, one and zero directions
along the light cone. 

\subsection{Light cone branes: $(+, -, m, n)$ branes}

First let us consider Dp-brane embeddings 
whose longitudinal directions include the light cone, whose
transverse positions are (arbitrary) constants and which carry
no worldvolume flux. Following the
discussion around (\ref{zfeq}) one can show that 
(almost) any Dp-brane embedding, with 
arbitrary $p$ and constant transverse scalars will satisfy 
the D-brane equations of motion. 

Suppose the Dp-brane longitudinal to the light cone 
has $m$ longitudinal directions
amongst the $y^a$, labelled by $(a_{1}..a_{m})$ and 
$n$ longitudinal directions amongst the $z^a$, labelled
by $(b_1... b_n)$; for convenience
of notation we will call this an $(+, -, m,n)$ Dp-brane. 
Then the allowed constant embeddings can be summarised as 
\bea
D1 & : & \hspace{10mm} (+, -, 0,0) \nonumber \\
D3 & : & \hspace{10mm} (+, -, 0,2) \hsp (+, -, 1,1) \hsp (+, -, 2,0) \\
D5 & : & \hspace{10mm} (+, -, 1,3) \hsp (+, -, 2,2) \hsp (+, -, 3,1) 
\nonumber \\
D7 & : & \hspace{10mm} (+, -, 2,4) \hsp (+, -, 3,3) \hsp (+, -, 4,2)
 \nonumber \\
D9 & : & \hspace{10mm} (+, -, 4,4) \nonumber
\eea
where in each case the transverse positions are arbitrary. In each
case the induced worldvolume metric is a pp-wave, as for
the $(+,-,3,1)$ brane discussed previously. Note that the
D9-brane fills the entire spacetime. 

The only possibilities not allowed in this table 
are $(+, -, 4,0)$ and $(+, -, 0,4)$ D5-branes. This is because
in these cases there is a non-zero WZ current which acts as a source
for the gauge field on the worldvolume, and so it is not consistent
to set this gauge field to zero. 
We will return to these exceptional cases below: they are directly
related to the baryon vertex \cite{Wit2} in the dual theory.   

All of the branes in the table above with $m > 0$ originate
as $AdS_{m+1} \times S^{n+1}$ branes. The exceptional cases
of $(+,-,0,0)$ and $(+,-,0,2)$ cannot originate from $AdS$
embeddings; instead these branes extend along the time direction
in $AdS$ and a maximal sphere in the $S^5$. The explicit forms
of these embeddings are most easily found in global coordinates:
write the $AdS_5 \times S^5$ metric as 
\be
ds^2 = R^2 [\cosh^2 \r d\t^2 + d\r^2 + \sinh^2 \r d\bar{\Omega}_3^2]
+ R^2 [d\th^2 + \cos^2 \th d\psi^2 + \sin^2 \th d\Omega_3^2].
\ee
The $(+,-,0,0)$ brane originates from a D1-brane extending 
along $(\t,  \psi)$ and located at $\r=0, \th=0$ and at arbitrary 
positions in $S^3$ and $\bar{S}^3$.  The brane has topology 
$R^1 \times S^1$ and the Penrose limit is taken by boosting
along this $S^1$. The $(+,-,0,2)$ brane originates from a D3-brane 
wrapping $(\t, S^3)$ and located at $\r=0$, $\th=\pi/2$, at
arbitrary $\psi$ and arbitrary position in $\bar{S}^3$. The brane has topology 
$R^1 \times S^3$ and the Penrose limit is taken by boosting
along an $S^1$ contained in the $S^3$. 
Notice that the D3-brane solution does not couple 
to the background RR-flux. When one rotates the brane about
a circle in the $S^5$ transverse to the wrapped $S^3$, the brane
will couple to the background flux and becomes the giant graviton,
which we will discuss in \S 8.4.

\bigskip

Now consider the supersymmetry of the brane embeddings into
the pp-wave background. The kappa symmetry projector is
\be
\Gamma =  \g_{+ -a_1..a_m b_1 ..b_n} K^{\frac{p+1}{2}} I \equiv
Q K^{\frac{p+1}{2}} I.
\ee
Recall that $K$ acts by complex conjugation, $I$ by multiplication by $-i$ and
note that $Q^2 = (-1)^{\frac{p-1}{2}}$. Following a similar analysis
as in section \ref{ppbr}, and using the fact that
the kappa symmetry projection should hold at all values of $x^+$
(since it is a worlvolume coordinate), one finds that the kappa symmetry 
projection yields three equations: the first coming from terms proportional
to $\cos^2 \half x^+$, the second from terms proportional to 
$\sin \half x^+ \cos \half x^+$, and the third from terms proportional to 
$\sin^2 \half x^+$. The analogous equations in section \ref{ppbr} 
are given in (\ref{spc1}). The first and third equations imply
\be \label{cond1}
\l = Q \eta; \qquad (Q P + (-)^{\frac{p-1}{2}} P Q ) \eta = 0; 
\ee  
whereas the second one yields
\be \label{cond}
(\g_{1234} + \g_{5678}) [(-1)^\delta Q \l + \eta] =0.
\ee  
where $\d$ is one if $(m,n)$ are both odd and
is zero if $(m,n)$ are both even. Multiplying (\ref{cond}) by 
$\g_{+-1234}$ and using the negative chirality of the 
$\eta$ and $\l$ spinors one obtains,
\be \label{cond2}
[(-1)^{\delta +{(p-1) \over 2}} +1] \g_{-} \g_+ \eta   =0,
\ee
where we have also used the first condition in (\ref{cond1}) 
and the fact that $Q^2=(-1)^{\frac{p-1}{2}}$.

The first condition in (\ref{cond1})
breaks the supersymmetry by one half by relating real and imaginary 
parts of the constant spinors. The second condition in (\ref{cond1})
will not impose further constraints provided that $Q$ and $P$
anticommute for $p = 1,5$ and commute for $p = 3, 7$. Let $P_{wv}$ be 
the part of $P$ that depends only on the worldvolume coordinates
and $P_{tr}$ the part that depends only on the transverse coordinates.
Notice that $P=P_{tr}$ in the case of D1-$(+,-,0,0)$ brane.
Explicit computation yields
\be \label{comm}
Q P_{wv} + (-)^{(\d+1)} P_{wv} Q =0, \qquad 
Q P_{tr} + (-)^\d P_{tr} Q =0
\ee
It follows that the second condition will be satisfied 
provided the brane is located at the origin of transverse space,
so that $P_{tr}=0$, and that
\be \label{cond3}
(-)^\d = (-)^{(p+1)/2} 
\ee
holds. The latter condition implies that (\ref{cond2}) is also satisfied. This 
yields the following set of 1/2 supersymmetric D-branes:
\bea
D3 & : & \hspace{10mm} (+, -, 0,2) \hsp (+, -, 2,0) \nonumber \\
D5 & : & \hspace{10mm} (+, -, 1,3) \hsp (+, -, 3,1) 
\\
D7 & : & \hspace{10mm} (+, -, 2,4) \hsp (+, -, 4,2)
 \nonumber
\eea
This agrees with the analysis 
of open strings in the pp-wave background reported in
\cite{dabholkar}.

If the branes are not located at the origin of transverse space 
one can satisfy the second condition in (\ref{cond1}) by imposing 
\be \label{g-}
\g_- \l = \g_- \eta =0
\ee
(except for the D1-brane that we discuss below).
This still leaves (\ref{cond2}) to be satisfied. If the 
index splitting is such that (\ref{cond3}) holds then 
the configuration preserves 1/4 of supersymmetry; otherwise
one needs to impose
\be \label{g+-}
\g_- \g_+ \eta=0
\ee
which together with (\ref{g-}) implies that $\eta=\l=0$, i.e.
the brane does not preserve any supersymmetry.\footnote{
We thank Marija Zamaklar and Pascal Bain for communication 
about this point and for pointing out an error in the original
version of this paper.}

The D1-$(+,-,0,0)$ brane is exceptional in that  $P=P_{tr}$. 
It follows from (\ref{comm}) that the  second condition in (\ref{cond1}) 
is satisfied automatically. The condition (\ref{cond3}) however
does not hold, so one should impose (\ref{g+-}).
Since in this case we do not impose (\ref{g-}),  
(\ref{g+-}) merely breaks the supersymmetry to one
quarter.

In summary, all configurations in the previous table preserve 
one half of supersymmetry when located at the origin of the transverse
space and one quarter when located at arbitrary positions,
the D1-$(+,-,0,0)$ brane preserves one quarter of supersymmetry
irrespectively of its location in transverse space, and all 
other embeddings listed at the beginning of this section do not 
preserve any supersymmetry.

\bigskip

Just as for the $(+, -, 3,1)$ branes discussed previously, it
is likely that switching on specific constant worldvolume fluxes may allow
us to move the branes away from the origin in the transverse
directions whilst preserving one half supersymmetry. It is 
certainly true that, as in the previous analysis, one can still 
satisfy the
field equations with constant transverse scalars if one switches
on constant fluxes $F_{+ a}$ on the worldvolume.
We have not explored in detail under what conditions such 
embeddings with constant fluxes are supersymmetric.

Before leaving the lightcone branes, let us briefly discuss
the exceptional case of $(+,-,4,0)$ branes ($(+,-,0,4)$ 
follow by exchanging $y^a$ with $z^a$). Recall that in this
case the pulled back
RR flux acts as a source for worldvolume flux. One can
verify that the field equations are satisfied for (arbitrary) 
constant transverse scalars provided that one switches on a 
worldvolume flux satisfying
\be
F = f_{a} dx^{+} \wedge dy^a; \hspace{10mm} 
\sum_{a=1}^{4} \partial_{a} f_{a} = 4.
\ee
So although a ``constant'' $(+,-,4,0)$ embedding does not
exist there is still a simple $(+,-,4,0)$ embedding with worldvolume
flux. 

The next question is whether this embedding preserves
supersymmetry: the analysis follows closely that given for 
$(+,-,3,1)$ branes with flux. One gets (\ref{spc1})
with $P$ is as in (\ref{zsc1}) but now
\be
Q = \g_{+- 1234}; \hsp R = \ep^{abcd} f_{a} \g_{-bcd}.
\ee
The first and third condition in (\ref{spc1})
reduce to those given in (\ref{cnc0}), namely
\be
\lambda = Q \eta; \hsp \{P, Q \} \eta = - Q R \eta; \hsp [Q,R] \eta = 0, 
\label{c40}
\ee
The second condition in (\ref{spc1}) together with the first 
equation above implies
\be \label{secd}
\g_{-} \g_+ \eta = 0.
\ee
 Given a general solution for $f_a$, we find that since
terms in $P$ involving the worldvolume coordinates commute with 
$Q$ the second condition in (\ref{c40}) can only be satisfied 
everywhere on the worldvolume provided that we impose 
the projection $\g_{-} \eta = 0$. This together with (\ref{secd})
implies that all supersymmetry is broken.

However, when $f_a = y^a$, the supersymmetry is enhanced
to one quarter, even when the brane is not located at the origin
in the transverse space. This is because terms in $P$ involving
the transverse coordinates automatically anticommute with $Q$
and drop out of the second condition. The terms
in $P$ involving the worldvolume coordinates, $P_{wv}$, still
commute with $Q$ but
one now finds that $R= - 2 P_{wv}$ and hence the second and the 
last conditions in (\ref{c40}) are satisfied even without
imposing a further projection on $\eta$. It follows that 
this embedding preserves one quarter of supesymmetry irrespective
of the location of the brane.

%Following the logic of this paper, that half supersymmetric
%embeddings in $AdS$ map to half supersymmetric embeddings
%in the pp-wave background, we postulate that this indicates
%there are corresponding
%half supersymmetric embeddings in $AdS$: D5-branes wrapping
%the $S^5$ with flux on the sphere or D5-branes wrapping
%$AdS_5 \times S^1$ with flux on the $AdS_5$. We leave these for future 
%investigation. 

Note also that there may also be half supersymmetric branes 
whose longitudinal directions are $(x^{+},x^{-},z^a)$ for which
the transverse scalar $r = \sqrt{y^a y^a}$ is {\it not}
constant. These would arise from the Penrose limits of the branes 
dual to the baryon vertex found in \cite{Craps}.

\subsection{Instantonic branes: $(m,n)$ branes}

Now let us consider so-called instantonic branes which
are transverse to the light cone directions. Such branes
would arise from the Penrose limit of Euclidean branes in
$AdS$. Let us again
take the transverse scalars to be constant and assume there
is no worldvolume flux. Solving the D-brane equations of motion
(\ref{finfe}) (with $\sqrt{-M}$ replaced by $\sqrt{M}$ due
to Wick rotation on the woldvolume theory),
and employing the notation $(m,n)$ we find the following
possibilities
\bea
D1 & : & \hspace{10mm} (0,2) \hsp (1,1) \hsp (2,0 )\nonumber \\
D3 & : & \hspace{10mm} (1,3) \hsp (2,2) \hsp (3,1) \\
D5 & : & \hspace{10mm} (2,4) \hsp (3,3) \hsp (4,2) 
\nonumber \\
D7 & : & \hspace{10mm} (4,4) \nonumber.
\eea
Notice that since we consider only embeddings 
with no worldvolume fluxes and no couplings to background fluxes, 
the computation is insensitive to the signature of spacetime.
In each case the induced worldvolume metric is just the flat Euclidean 
metric.
We must leave out the $(4,0)$ and $(0,4)$ D3-branes for reasons
akin above: the background RR flux acts as a source for
worldvolume scalars. The analysis in the previous section
suggests that there are such half-supersymmetric D3-branes.

Now consider the supersymmetry of these embeddings. We run
into a problem when trying to use the kappa symmetry projection:
the Euclideanised kappa symmetry projector has not (as far
as we know) been constructed in the literature. To construct
such a projector one may Wick rotate the worldvolume 
theory using the results in 
\cite{Nicolai:1978vc,vanNieuwenhuizen:1996tv,Waldron:1997re}
and then demand that the action is invariant under the 
new Euclideanised kappa symmetry. A similar approach 
was taken in \cite{Ooguri:1998pf} when studying the 
supersymmetry of D-instantons.

Such instantonic branes were constructed as closed string
boundary states \cite{billo} and
it was found that the following branes preserve half the supersymmetry
when located at the origin in the transverse spatial coordinates
and at arbitrary positions on the lightcone
\be
D1  :  (0,2), (2,0) \hsp D3  : (1,3), (3,1) \hsp
D5 : (2,4), (4,2). 
\ee
All other possibilities preserve one quarter supersymmetry. It
would be interesting to investigate corresponding (Euclidean)
brane embeddings in $AdS_5 \times S^5$ which are as yet unknown. These
correspond to near horizon limits of Euclidean D-branes intersecting
with the background D3-branes and are likely to have topologies
of the form $H^{m} \times S^{n}$. 

\subsection{Branes along one lightcone direction: $(+_{\g},m,n)$ branes} 

Although in string theory in light cone gauge only branes
totally transverse or totally longitudinal to the light cone are accessible, 
branes along only one light cone direction could also 
exist. Let us take the brane to lie along $x^{-} = \g x^{+}$.
Such branes are certainly 
physically interesting objects, arising from the Penrose limit
of branes in $AdS_5 \times S^5$ which are rotating in the circle
direction, at a speed implied by $\g$. Branes which
rotate at the speed of light in the direction of the boost
have $\g = 0$ and lie along
the $x^{+}$ direction at some constant value of $x^{-}$. These
could be called $(+,m,n)$ branes, following the previous notation.

Note that branes for which $\g \rightarrow \infty$,
in other words $(-,m,n)$ branes for which $x^{+}$ is constant, and 
for which the other transverse scalars
are constants would have degenerate worldvolume metrics, since
$g_{--} = 0$, and hence are not admissible solutions. All other
possibilities in the range $0 \le \g < \infty$ could in principle be
realized and would correspond to branes rotating at speeds depending
on $\g$ along the boosted circle. 

\bigskip

To find such branes from our field equations, let us take
an ansatz in which the brane extends along $(x^{+}, m, n)$
with the lightcone transverse scalar being 
$x^{-} = x^{-}_{0} + \g x^{+}$ and all the other 
transverse scalars constant. The D-brane field 
equations permit solutions {\it only} when the transverse scalars are 
zero and in the following cases
\bea
D1 & : & \hspace{10mm} (+_{\g},0,1) \hsp (+_{\g},1,0) \nonumber \\
D3 & : & \hspace{10mm} (+_{\g},1,2) \hsp (+_{\g},2,1) \\
D5 & : & \hspace{10mm} (+_{\g},2,3) \hsp (+_{\g},3,2) \nonumber \\
D7 & : & \hspace{10mm} (+_{\g},3,4) \hsp (+_{\g},4,3), \nonumber
\eea
where we use as shorthand $+_{\g}$ to indicate the lightcone
direction wrapped by the brane.
The value of $\g$ is left arbitrary by the field equations
though as mentioned above the solution with $x^{+}$ strictly constant is
not admissible because the induced metric degenerates. The value of
$x^{-}_{0}$ is also an an arbitrary integration constant.
These branes all have induced metrics:
\be
ds^2 = - (r^2 + 4 \g) (dx^+)^2 + dr \cdot dr,
\ee
where $r^2 = \sum_{a=1}^{m} (y^a)^2 + \sum_{a=1}^{n} (z^a)^2$.
As in the previous
discussions, several possibilities are missing from this table, 
because of the coupling to the RR flux. We will consider below 
$(+_{\g},0,3)$ and $(+_{\g},3,0)$ branes, for which the RR flux couples to
a transverse scalar. $(+_{\g},1,4)$ and $(+_{\g},4,1)$ branes are
also missing from this table, because the RR flux again acts
as a source for the worldvolume flux, and we do not consider them 
here. 

\bigskip

Starting from our
$AdS_{m+1} \times S^{n+1}$ embeddings, if the boost circle is
not contained in the wrapped sphere, we find that the Penrose limit 
of the brane gives the $(+_{\g},m,(n+1))$ branes found 
above but with $\g \rightarrow \infty$.

The simplest case to consider explicitly is the $AdS_2$ brane. 
This can be
embedded in global coordinates by taking the brane to extend along
$\tau$ and $\rho$ at $\chi = \psi = 0$, $\phi=const$ 
and at constant position in
the $S^5$. In particular this means that $\th_5$ is constant and
hence when we change to the lightcone coordinates
\be
dx^{+} = \frac{dx^{-}}{R^2}
\ee
on the brane. Now we take the limit $R \gg 1$, giving an induced
metric on the brane
\be
ds_2^2 =  - (r^2 + 4 R^2) (dx^+)^2 + dr^2,
\ee
with $x^{-} = x^{-}_{0} + R^2 x^{+}$. This indeed reproduces
the metric for the $(+_{\g},1,0)$ brane found above, though in 
the Penrose limit we would need
to take $R \rightarrow \infty$. In this limit the brane lies 
strictly parallel to the $x^{-}$ lightcone direction; then
the induced metric must be degenerate (since $dx^{+} = 0$) and furthermore
the brane has infinite energy as measured by the lightcone Hamiltonian.  
Note that the infinite boost means that branes which are static with 
respect to the boost will always be mapped to branes along $x^{-}$ 
in the pp-wave background.  

\bigskip

Now let us consider the supersymmetry of these embeddings. 
Whenever we take the Penrose limit of a (static) $AdS_{m+1} \times 
S^{n+1}$ branes using a circle transverse to the brane, the
resulting brane will have a degenerate metric and is not an admissible
solution. Even branes which were originally half supersymmetric will 
not appear as admissible embeddings in the pp-wave background. 

Branes in the pp-wave background with a finite value of $\g$ originate
from branes which were rotating close to the speed of light about
the boosted circle. Generically rotating or boosted 
branes are not supersymmetric and we will find this is true here:
virtually none of the $(+_{\g},m,n)$ branes preserve any
supersymmetry even when located at the origin in transverse 
coordinates. The only exceptions we have found 
are the $(+,0,1)$ and $(+,1,0)$
branes as well as the (exceptional) $(+,3,0)$ branes discussed
in the next section. 

The kappa symmetry projector for the $(+_{\g},1,0)$ branes is 
\be
\G = \frac{1}{(r^2 + 4 \g)^{\frac{1}{2}}} 
( \g_{+} - \half (r^2 + 4 \g) \g_{-} ) 
\g_{1} K I \equiv (Q_{+} + Q_{-}) K I, 
\ee
where in the latter expression we introduce $Q_{\pm}$
which depend on $\g_{\pm}$ respectively. Let us
write the projection as 
\be
\G (1 + i P) \psi(x^{+}) = -(1 + i P) \psi(x^{+}),
\ee
where $P$ is given in (\ref{zsc1}) and 
\be \label{psi}
\psi(x^+) = (\cos(\half x^{+}) - i \sin (\half x^{+}) \g_{1234})
(\cos(\half x^{+}) - i \sin \half x^{+} \g_{5678}) (\lambda + i \eta)
\ee
is a spinor of negative chirality. Using the nilpotence
of $\g_{-}$ this reduces to 
\be
i (Q_{+}   + Q_{-}  - i Q_{+} P ) \psi^{\ast}(x^{+}) = (1+ i P) 
\psi(x^{+}). \label{gcs}
\ee
Now this condition must hold everywhere on the worldvolume, namely
for all $r$. Writing $P$ on the worldvolume as
\be
P = -\half r \g_{-234} 
\ee
(all other terms vanish when the transverse scalars are zero) we find
that necessary and sufficient conditions for (\ref{gcs}) to hold are
that $\g = 0$ and in addition
\be
\g_{+} \psi(x^{+}) = 0; \hspace{10mm} \g_{1234} \psi(x^{+})^{\ast}
= \psi(x^{+}),
\ee
which can be satisfied by 
\be
\g_{1234} \l = \l; \hspace{10mm} \g_{1234} \eta = - \eta; 
\hspace{10mm} \g_{+} \l = \g_{+} \eta = 0.
\ee
The projections break the supersymmetry to one quarter maximal. 

\bigskip

Now let us consider the origins of these 1/4 supersymmetric branes
in $AdS_5 \times S^5$. The $(+,1,0)$ branes come from the following
embedding: use global coordinates for $AdS_5$ and
extend the string along $(\rho, \tau)$ with the other $AdS_5$ 
coordinates $\chi$, $\psi$ and $\phi$ fixed.
For the transverse positions
in the $S^5$, take all $\th_a = \half \pi$, except $\th_5 = \tau$.
The induced worldvolume metric is
\be
ds^2 = R^2 ( - \sinh^2 \rho d\tau^2 + d \rho^2),
\ee
which is an $AdS_2$ metric. Since the brane rotates at the speed
of light around the $\th_5$ circle, we will refer to this brane
as a rotating $AdS_2$ string.

To find the origin of the $(+,0,1)$ brane it is convenient to
first introduce the following coordinates on the $S^5$:
\be
ds^2 = (d\th^2 + \sin^2 \th d\Omega_3^2 + \cos^2 \th d\th_5^2).
\ee
Then put the D1-brane at $\rho = 0 $ in the $AdS_5$ along
$\tau$ and along $\th$ in the $S^5$ at fixed position
on the $S^3$ with $\th_5 = \tau$. 
The induced metric in this case is just
\be
ds^2 = R^2 ( -\sin^2 \th d\tau^2 + d\th^2),
\ee
which we will refer to as a rotating $dS_2$ string (the induced
metric is de Sitter).

Note that there is no reason to suppose that these embeddings should
be supersymmetric {\it a priori} since they do not come
from any standard intersecting brane system. Since we find the 
corresponding pp-wave embedding is only one quarter supersymmetric
these embeddings in $AdS$ preserve
at most one quarter supersymmetry and possibly none. It would
be interesting to check their supersymmetry explicitly. 

These rotating D-strings are expected to correspond to magnetic 
monopoles from the gauge theory point of view.
Since the branes are rotating about a circle, the states also carry 
charge with respect to the corresponding $SO(2)$ R symmetry. 
Note that the static $AdS_2$ brane considered 
above also has a mass of the same
order but it does not carry any charge with respect to the
R symmetry; it must therefore correspond to a monopole with no R 
charge, which is why its lightcone mass diverges in the Penrose
limit. Although the masses of the rotating D-strings 
also diverge in the Penrose limit, 
they can still have finite lightcone Hamiltonian because the R-charge
can cancel with the mass term.

\subsection{An exceptional case: the $(+,0,3)$ branes and giant
  gravitons in $AdS$}

Finally let us consider the $(+,3,0)$ and $(+,0,3)$ branes which
are missing from the above classification, because of their coupling
of the RR flux to the worldvolume scalars. It turns out that
there is a simple D3-brane embedding along one direction of
the lightcone. Let us treat the $(+,3,0)$ brane; again the other
case follows by simply exchanging of the $y^a$ and $z^a$ coordinates. 

Take the longitudinal directions to be
$(x^{+}, \Omega_3$), where we now use polar coordinates $(r,\Omega_3)$ 
to describe the $R^4$ parametrised by $y^a$. The field equations 
(\ref{finfe})
for the (constant) transverse scalars $r$ and $z^a$ are respectively
\bea
4 r^3 &=& r^2 (4 r^2 + 3 z^a z^a) (r^2 + z^a z^a)^{-\frac{1}{2}}; \\
0 &=& r^3 z^a (r^2 + z^a z^a)^{-\frac{1}{2}}, \nn
\eea
where in the first equation there is a WZ source term arising
from the background flux $f_{+r \Omega_3} = 4 r^3$. These equations
are manifestly 
satisfied provided that $y^a y^a = r^2$ is an arbitrary 
constant (the brane wraps an $S^3$ in this $R^4$) and the other transverse 
scalars $z^a$ are zero. Note that the lightcone transverse 
scalar $x^{-}$ is also constant, so that the brane rotates at the
speed of light along the boosted circle. 
    
The induced worldvolume metric on the brane is just
\be
ds^2 = - r^2 (dx^+)^2 + r^2 d\Omega_{3}^2,
\ee
which is an Einstein universe of radius $r$.  Thus the effect
of the coupling to the flux is that the spatial sections
of the $(+,3,0)$ brane are spherical rather than flat. 
Note that $r$ is not necessarily zero: surprisingly the field
equations allow the brane to have a finite radius, although one
would naively expect it to shrink. This stabilisation results
from the coupling to the background RR flux.

To check supersymmetry, note first that the projection operator is 
\be
\Gamma = i r^{-1} (\g_{+} - \half r^2 \g_{-}) \g_{234} = i Q,
\ee
where $Q^2 = -1$. Then note that on the worldvolume $z^a = 0$
with $r$ constant we can rewrite $P$ in (\ref{zsc1}) as
\be
P = - \half r \g_{-} \g_{234}
\ee
where we relate gamma matrices in polar and Cartesian frames by 
$\g_{r} = \g_{1}$, $\g_{\th_1} = r \g_{2}$ and so on. Now
the kappa symmetry projection is 
\be
i Q ( 1 + i P) \psi(x^{+}) =  ( 1 + i P) \psi(x^{+}),
\ee
where $\psi(x^+)$ is given in (\ref{psi}). Using the nilpotence of $\g{-}$ 
this reduces to
\be
i \tilde{Q} (1 + iP) \psi(x^{+}) =  \psi(x^{+}) \label{czn1}
\ee
where $\tilde{Q} = r^{-1} \g_{+234}$. This can be
solved for any $r$ at all values of the worldvolume
coordinate $x^{+}$ by taking
\be
\g_{+} \psi(x^+) = 0. \label{czn2}
\ee
This projects out half of the $\lambda$ and $\eta$ spinors
contained in $\psi(x^{+})$ and kills the first term in (\ref{czn1}).
Noting that 
\be
- \tilde{Q} P \psi(x^+) =  \half \g_{+} \g_{-} \psi(x^+) 
= \psi(x^+),
\ee
where we use (\ref{czn2}) and $\{\g_{+},\g_{-} \} = 2$, 
we see that (\ref{czn1}) is indeed satisfied.
Thus, perhaps somewhat surprisingly, the brane embeddings preserve
one half supersymmetry for any value of $r$. 
In a slight abuse of notation, these could be called $(+,3,0)$
branes (the abuse is that the branes extend along an $S^3$ and not 
an $R^3$ of the $R^4$). Embeddings closely related to these
in which $r$ is a function of $x^{+}$ will be the missing
$(4,0)$ instantonic branes, though the embeddings are no longer constant:
indeed they must be explicitly time dependent. 

\bigskip

Now we have argued that branes which preserve one half
supersymmetry will originate from supersymmetric configurations in 
$AdS_5 \times S^5$. So this analysis indicates that there is
are stable rotating D3-branes in $AdS_5 \times S^5$. These
branes correspond to giant gravitons \cite{Len,Itz,My1}.
Again the explicit form of the embedding is easiest to find
using global coordinates. The $(+,3,0)$ embeddings originate
from branes which extend along the directions $(\tau,\chi, \psi,\phi)$
in the $AdS_5$ at {\it arbitrary} radius $\rho_{0}$, with $\th_5 = \tau$
parametrising the rotation in the $S^5$ and all other angular
coordinates in the $S^5$ being $\pi/2$. Then the induced worldvolume
metric is just
\be
ds^2 = R^2 \sinh^2 \rho_{0} [- (d\tau)^2 + (d\chi^2 + 
\cos^2 \chi d\psi^2 + \sin^2 \chi d\phi^2) ],
\ee
which is an Einstein universe, just as for the Penrose limit
of the brane. Note that if the brane was not rotating it
would be forced towards $\rho \rightarrow 0$; the rotation
stabilises it at finite $\rho_{0}$. It was shown in \cite{Itz,My1}
that this embedding is half supersymmetric.

To find the origin of the $(+,0,3)$ embedding in $AdS_5 \times 
S^5$ it is easiest to use global coordinates for the $AdS$ and
introduce coordinates on the $S^5$
\be
ds^2 = R^2 [d\th^2 + \sin^2 \th d\Omega_3^2 + \cos^2 \th d\th_5^2].
\ee
The following ansatz satisfies the field equations: 
the brane wraps the $S^3$ in this $S^5$ and rotates at the
speed of light in the $\th_5$ direction, so that $\tau = \th_5$,
with all other transverse scalars (the other coordinates on the
$AdS_5$ and $\th$) constant. Then the field equations are
satisfied with $\rho = 0$ but {\it arbitrary}
$\th = \th_{0}$. The induced metric on the brane is then
\be
ds^2 = R^2 \sin^2 \th_{0} [ -(d\tau)^2 + d\Omega_3^2],
\ee 
which is again an Einstein universe. This embedding was
also discussed in \cite{Itz,My1} and shown to be half
supersymmetric.

Now the giant graviton is, as we mentioned earlier, closely
related to the $R \times S^3$ embeddings: the giant graviton
is obtained by boosting this brane along a transverse $S^1$
in the $S^5$. We have discussed here taking the Penrose
limit by boosting along the transverse $S^1$ about which
the brane is rotating. One could also take the Penrose limit
using a circle in the longitudinal $S^3$: this would lead
to a $(+,-,0,2)$ brane which rotates in an $R^2$ transverse
to the plane. It would thus appear in
the spectrum of our ``static'' $(+,-,0,2)$ brane
as an excited state carrying
angular momentum.

\bigskip

Finally, let us say that 
the whole story of D-brane embeddings in pp-wave 
and $AdS \times S$ backgrounds 
seems to be very rich and deserves much more investigation. We hope
to have pointed out a number of avenues to pursue: we have classified
all constant brane embeddings in the pp-wave background 
but embeddings arising from the baryon vertex branes, those 
with (non)constant fluxes on the worldvolume and rotating branes
should be explored much further. 
By essentially inverting the Penrose
limit, these branes in the pp-wave background will 
lead us to previously unknown brane embeddings 
in $AdS_5 \times S^5$, beyond the ones discussed here.
Such embeddings could in turn lead us to new results in the
corresponding dual field theories. 

\section{D-branes from gauge theory} \label{gauD}

The pp-wave limit of $AdS_5 \times S^5$ 
has attracted interest principally because one can directly construct 
the light-cone string theory in the pp-wave background from gauge 
theory \cite{BMN}. The authors of \cite{BMN} constructed the 
light-cone closed string states using specific operators of the 
$d=4 \ \cn=4$ SYM. A natural question is how to construct the D-brane
states we found here using gauge theory. Since D-branes capture
non-perturbative aspects of string theory this question is
of paramount interest. We will present here a construction of
all (longitudinal) D-branes appearing in the Table 1 of the introduction.
As this paper was being typed, two very interesting papers 
appeared, \cite{BMN2} and \cite{Lee:2002cu}, where the 
construction we outline below was carried out in detail 
for the case of the D7-$(+,-,4,2)$ brane and the D5-$(+,-,3,1)$ brane,
respectively. 

As already discussed, there is a one to one correspondence 
between supersymmetric intersections $(Dq \perp D3)$,
supersymmetric $AdS$ D-branes in $AdS_5 \times S^5$ and 
Dq-branes along the light cone in the pp-wave background. We 
will make use of all three to construct the latter Dq-branes from 
the gauge theory.

Firstly, following the arguments in \cite{KR,ooguri1}, one expects 
an AdS/dCFT duality for all cases appearing in table 1.
That is, we expect a duality between the bulk 
theory on $AdS_5 \times S^5$ together with a Dq-brane
probe and the boundary theory $d=4 \ \cn=4$ SYM with a defect CFT 
on the boundary of the $AdS$-embedding. The case of $(3|D7 \perp D3)$
(leading to the D7-$(+,-,4,2)$ brane in the pp-wave background)
is somewhat special in that the ``defect CFT'' is actually
of ``co-dimension zero'', i.e. the D3-branes lie entirely 
on the worldvolume of the D7-brane and the fields coming from 3-7 
and 7-3 strings are localized on the worldvolume of the D3-brane.
As is well known, one needs to include an 
orientifold plane in this case. The corresponding duality is well 
understood \cite{Fayyazuddin:1998fb}, \cite{Aharony:1998xz} and we 
refer to these papers for further details. 

In the limit discussed in \cite{ooguri1}, the bulk theory captures 
the physics of the closed strings and of the q-q open strings,
whilst the boundary theory captures that of the 3-3, 
3-q and q-3 open strings. The boundary action can be 
obtained from the action of Dq-D3 system upon taking the 
near-horizon limit discussed in \cite{ooguri1}.
In particular, we note that the 3-q and q-3 open strings will
give rise to hypermultiplets in the  fundamental of $SU(N)$. 
These defect fields interact amongst themselves and with the 
restriction of the $\cn=4$ vector multiplet to the defect.
As argued in \cite{ooguri1}, the defect theory will
capture holographically the physics of the q-q open strings.

To make the duality precise one needs to develop a dictionary 
between bulk fields and boundary operators. In particular, the 
worldvolume fields of the Dq-brane reduced over the wrapped 
sphere should match with (certain) operators of the defect theory. 
This has been done explicitly for the $(2|D5 \perp D3)$ and 
the $(3|D7 \perp D3)$ cases in \cite{ooguri1} (as discussed
in \S\ref{holdual}) and in
\cite{Aharony:1998xz}, respectively. Notice that 
all configurations in the top part of Table 1 
have four Neumann-Dirichlet (ND) boundary conditions
and the ones in the bottom part have eight.
This means that all AdS/dCFT pairs of the top/bottom part of the table
are formally connected by T-duality (formally because
one would have to wrap the branes on tori to perform 
the T-duality, an operation which in general does not 
commute with the near horizon limit). This implies 
that a consistent field/operator matching is guaranteed.
 
Having argued the existence of an AdS/dCFT in each case, we 
now take the pp-wave limit of these configurations.
The construction of the closed string spectrum
proceeds as in \cite{BMN} and we shall not repeat it
here; in what follows we use the notation of \cite{BMN}
when referring to the closed string sector.
We only mention that the closed string vacuum is 
constructed from the $Z$'s and the oscillators from 
insertions of $D_i Z$ and $\phi_i$,
where $Z=X^4+i X^5, \bar{Z}, \phi^i=X^{i-5}$,
and $X^i,i=4,..,9,$ are the six scalars of $\cn=4$ SYM.

As we have seen, AdS embeddings are mapped to  
D-branes along the light cone in the pp-wave background, 
suggesting that the 
D-brane states can be constructed using defect fields.
In particular, to construct the open string states one needs to have 
fields in the fundamental. For the $ND = 4$ cases these are supplied 
by the hypermultiplets localized on the defect. To construct
the appropriate states we will
need to know the $J$ charge of the hypermultiplets with respect to
$SO(2)$ associated to the circle along which we boost.
The easiest way to compute this is to go back to the original
Dq-D3 system and find the $J$ charge by looking at the way the 
hypermultiplets are constructed from the fermionic zero modes of 
the q-3 and 3-q
strings. The computation is identical in all three cases, i.e.
D3, D5 and D7 branes and so it will suffice to consider 
just one case.
Since the D5 and D7 branes have appeared already 
in the literature let us discuss explicitly here the D3-brane.
The intersecting D3-D3' brane configuration is illustrated 
in the table below.
\begin{table}[h]
\begin{center}
\begin{tabular}{|c|c|c|c|c|c|c|c|c|c|c|}
\hline 
    & 0 & 1 & 2 & 3 & 4 & 5 & 6 & 7 & 8 & 9 \\ \hline 
D3' & N & N & N & N & D & D & D & D & D & D \\ \hline 
D3  & N & N & D & D & N & N & D & D & D & D \\ \hline 
\end{tabular} \caption{D3-brane intersection}
\end{center}
\end{table}
We remind our readers that to get to the D3 brane in the pp-wave
background from this configuration we will 
first need to go to the near-horizon
limit of the D3' branes, with the 45 directions of the D3 brane
extending along the radial direction in $AdS_5$ and along 
an $S^1$ in $S^5$. This produces the half supersymmetric 
$AdS_3 \times S^1$ embedding in $AdS_5 \times S^5$, as we have
discussed. We then take the pp-wave limit by boosting along this
same circle, which leads to the half supersymmetric 
D3-$(+,-,2,0)$ brane of the pp-wave background. 
Thus the $SO(2)$ that participates in the pp-wave limit 
acts as a rotation in the 4-5 plane. 

Now let us go back to the original intersecting D-brane system
to find the $R$-charges of the defect hypermultiplets.
This follows from an analysis exactly analogous to that 
given in \cite{Polchinski:rr}, where the 5-9 and 9-5 
strings of the D5-D9 system were studied. The massless 
states of the 3-3' system form a ``half-hypermultiplet''.
The scalars $q_i,i=1,2$, are in the fundamental of $SU(N)$ and transform
as $(1/2,1/2)$ and $(-1/2,-1/2)$ under rotations in the 45 and 23 
planes. Similarly, the 3'-3 strings yield yield another
``half-hypermultiplet'' containing two scalars $\bar{q}^i$
in the anti-fundamental of $SU(N)$ and transforming the same way as $q_i$ 
under rotations in the 45 and 23 planes. 
Thus we find that $q_1$ and $\bar{q}^1$ have $J=1/2$ and other two $J=-1/2$.
Both $q_i$ and $\bar{q}^i$ are singlets under the $SO(4)$ that rotates
the DD directions. Analogous analysis and results apply to the the D5 and 
D7 branes.

We now propose that the open string vacuum is all cases is given by
\be \label{openvac}
|0; p^+ \> \leftrightarrow  \bar{q}^1 Z^J q_1
\ee
Note that the $Z$ appearing in this formula denotes the restriction 
of $Z$ to the defect. This formula is schematic in the 
case of the D7-brane since we do not take into account the 
orientifold; a more precise formula for this case 
can be found in \cite{BMN2}. 

To compute the light-cone energy we need to know the dimension
$\D$ of the $q_i$ and $Z$. The dimension of $Z$ is always one,
and the dimension of $q_i$ depends on the spacetime dimension:
$\D_d=(d-2)/2$. Hence the light-cone energy of (\ref{openvac})
will vary from case to case,
\be \label{grounden}
D3: (\D-J)_0=-1, \qquad D5: (\D-J)_0=0, \qquad D7: (\D-J)_0=1.
\ee
One should compare this with the ground state energy 
of the corresponding D-brane. The number of 
bosonic massive zero modes is equal to $(p-1)$ for a $Dp$ brane,
and each of them contributes $1/2$ to the ground state energy.
In all three cases there are four massive fermionic 
zero modes, each contributing $-1/2$ to the ground state energy.
Combining these results we find exact agreement with (\ref{grounden})!

The discussion so far refers to the branes originating 
from $ND=4$ systems. Let us now discuss the $ND=8$ systems.
In this case the massless spectrum of the 
q-3 and 3-q strings consists of two fermions, $\chi$ and
$\bar{\chi}$, one in the fundamental and the other in the 
anti-fundamental of $SU(N) $\cite{Green:1996dd,Bachas:1997sc} . 
These originate from the R sector and are
singlets under rotations in the ND directions;
we hence conclude that both $\chi$ and
$\bar{\chi}$ have $J$ charge equal to zero.
We now propose that the open string vacuum is given by
\be \label{openvac1}
|0; p^+ \> \leftrightarrow \bar{\chi} Z^J \chi
\ee
Notice that the canonical dimension of $\chi$ 
in $d$ spacetime dimensions is $\D_\chi=(d-1)/2$ and
it hence follows that the light-cone energy of (\ref{openvac1}) 
in all cases is given exactly by (\ref{grounden})! 

One may now follow the discussion in \cite{BMN} and construct 
oscillators by insertions of $\phi^i$ and  
the covariant derivatives of $D_i Z$. We shall only 
discuss the bosonic oscillators but the fermionic ones 
can be constructed along similar lines. In the case of the 
open string states one would insert the restriction 
of these fields to the defect. To construct the non-zero modes 
one needs to add phases to the closed string states, as 
in \cite{BMN}, and cosines and sines for open strings with Neumann
and Dirichlet boundary conditions, respectively, as in 
\cite{BMN2,Lee:2002cu}. 

One still has to specify which operator should be inserted
to obtain a given oscillator. This follows from the form 
of the original intersection: one inserts a $D_i Z$
in the direction that was parallel to the $i$ 
worldvolume direction of the D3-brane in the 
original intersection, and $\phi^i$ in the directions 
that were parallel to the $\phi^i$ direction 
transverse to the D3-brane. These rules follow
from the decomposition of the vector multiplet
of the original D3-brane into other multiplets 
when restricted to the defect.

These rules apply equally to the $ND=4$ and $ND=8$ cases.
For illustrative purposes we present the case
of oscillators of the D3-$(+,-,2,0)$ brane,
\bea
&&a_n^{0 \dagger} |0; p^+ \> \leftrightarrow 
{1 \over \sqrt{J}} \sum_{l=0}^{J+1} {\sqrt{2}
\cos \left({n \pi l \over J} \right) \over N^{J/2+1}}
\bar{q}^1 Z^l (D_0 Z) Z^{J+1-l} q_1 \nonumber \\
&&a_n^{1 \dagger} |0; p^+ \> \leftrightarrow 
{1 \over \sqrt{J}} \sum_{l=0}^{J+1} 
{\sqrt{2} \cos \left({n \pi l \over J} \right) \over N^{J/2+1}}
\bar{q}^1 Z^l (D_1 Z) Z^{J+1-l} q_1 \nonumber \\
&&a_n^{2 \dagger} |0; p^+ \> \leftrightarrow 
{1 \over \sqrt{J}} \sum_{l=0}^{J+1}  
{\sqrt{2} \sin \left({n \pi l \over J} \right) \over N^{J/2+1}}
\bar{q}^1 Z^l (D_2 Z) Z^{J+1-l} q_1 \\
&&a_n^{3 \dagger} |0; p^+ \> \leftrightarrow 
{1 \over \sqrt{J}} \sum_{l=0}^{J+1} 
{\sqrt{2} \sin \left({n \pi l \over J} \right) \over N^{J/2+1}}
\bar{q}^1 Z^l (D_3 Z) Z^{J+1-l} q_1 \nonumber \\
&&a_n^{ {i+5} \dagger} |0; p^+ \> \leftrightarrow 
{1 \over \sqrt{J}} \sum_{l=0}^{J+1} 
{\sqrt{2} \sin \left({n \pi l \over J} \right) \over N^{J/2+1}}
\bar{q}^1 Z^l \phi^i Z^{J+1-l} q_1 \nonumber
\eea
Here $a^0_n$ and $a^1_n$ are oscillators of the worldvolume coordinates of the 
D3 brane other than the light-cone coordinates (the zero superscript 
does not mean that this is timelike coordinates), 
and $a^2_n, a^3_n, a^{i+5}_n$ are the oscillators of the 
six transverse directions. Notice that only the $a^0$ and $a^1$ 
have zero modes.
One can check that the assignments in \cite{BMN2} and \cite{Lee:2002cu}
follow from this rule.

One may proceed to compute the anomalous dimensions of these 
operators as in \cite{BMN,BMN2,Lee:2002cu}. For this one 
would need to know the precise form of the interactions 
between the boundary and defect fields. One may use 
T-duality to obtain these interactions from the interactions 
given in \cite{ooguri1}. This suggests that the arguments
in \cite{Lee:2002cu} carry over for this case as well.
Another avenue is to relate this system to the D1-D5 system. 
We leave a detailed investigation for future work.

We have presented the construction of all half
supersymmetric D-branes that are visible 
in the light-cone gauge and do not carry worldvolume
flux.  It would be interesting to incorporate the effects of the 
worldvolume flux in the construction 
and to understand how to construct the rotating 
D1 and D3 branes. We expect these to appear as non-perturbative 
states.

\section*{Acknowledgments} 
The research of KS is partially supported by
the National Science Foundation grant PHY-9802484. MT would like
to thank Princeton University for hospitality during the course
of this work. 

\appendix

\section{Conventions}

In \S 3 and onwards we use the following conventions for the type IIB 
field equations. Since the only non-zero fields are the metric
and the four-form potential we need only the Einstein equation which is
\be
R_{mn} = {\textstyle{1\over 96}}  f_{pqrsm} 
f^{pqrs}_{\hspace{6mm} n} \nonumber \\
\ee
where we must also 
impose the self-duality constraint on $f_{mnpqr}$. 
The five-form $f_{mnpqr}$ is 
defined in terms of the RR 4-form $C_{mnpq}$ as
\be
f_{mnpqr} = 5 \del_{[m} C_{npqr]} 
\ee
where here and elsewhere square brackets denote antisymmetrisation
with unit weight. Note that this normalisation of the RR field is
consistent with that appearing in the D-brane action (\ref{dbiact}).
With this truncation of the IIB equations 
the supersymmetry transformation for the dilatino
$\lambda$ is zero automatically and the gravitino $\psi_m$ 
variation is
\be
\delta \psi_{m} = (D_{m} \ep + \textstyle{i \over {1920}} \G^{pqrst} 
\G_{m} f_{pqrst} \ep). 
\ee
We use the following spinor conventions. We work with a mostly positive 
Lorentz metric $\eta_{mn}$ and Dirac $\g$-matrices obeying 
$\{ \g_m,\g_n \} = 2 \eta_{mn}$. The unit normalised matrices 
$\g_{a_1...a_n}$ are defined by
\be
\g_{a_1...a_n} = \g_{[a_1}...\g_{a_n]}. 
\ee
We reserve $\g$ for tangent space Dirac matrices and $\G$ for curved
space matrices.

\section{Kappa-symmetry and worldvolume flux}

We show in this appendix that the kappa symmetry projection 
(\ref{kap}) for the embedding with flux is related to the 
kappa symmetry projection for the embedding with no flux 
by a similarity transformation. This is a general property
of kappa symmetry projectors and is discussed 
in \cite{Bergshoeff:1997kr}. The novelty in our case is that 
the embedding itself depends explicitly on the flux. This introduces 
an additional dependence on the flux, through the dependence
of the Killing spinors on the spacetime coordinates.

The kappa symmetry projection in the case of no flux is 
given by
\be \label{noflux}
\G'=\g^{012489} KI
\ee
The similarity transformation that relates (\ref{noflux})
to the projector with flux is
\be \label{sim}
\G=e^{-a/2} \G' e^{a/2} 
\ee
where 
\be
a=\arctan(q) (\g^{89} K - \g^{34})
\ee
To prove (\ref{sim}) we first work out $e^{a/2}$. This can be
done by observing that $\g^{89} K$ and $\g^{34}$ square to 
minus one and commute with each other. After some algebra one
obtains,
\be
e^{a/2} = \cp'_- + {1 \over \sqrt{1+q^2}} (1 - q \g^{34}) \cp'_+
\ee
where 
\be
\cp'_{\pm} = \half (1 \pm \g^{3489} K).
\ee
Using both this result and the fact that $\G'$ 
commutes with $\cp'_{\pm}$ but anticommutes with $\g^{34}$
leads (after some more algebra) to (\ref{sim}).

Let $\e(x_0,q)$ the Killing spinor evaluated on the embedding surface
$x=x_0-q/u$. Then (\ref{sim}) implies that for any 
solution of the kappa symmetry projection with flux
\be \label{Gflux}
\G \e(x_0,q) = \e(x_0,q),
\ee
there is a solution of
\be \label{nofl1}
\G' \e'(x_0,q) = \e'(x_0,q)
\ee
with $\e'(x_0,q)=e^{a/2} \e(x_0,q)$. We stress that the 
equation (\ref{nofl1}) is distinct from the equation one 
would get by considering the zero-flux embeddings. In that 
case one would have
\be \label{nofl2}
\G' \e(x_0) = \e(x_0)
\ee
where $\e(x_0)$ is the target space Killing vector evaluated 
on the embedding surface $x=x_0$. Even though it is $\G'$
that features in both of (\ref{nofl1}) and (\ref{nofl2}),
the spinors involved are different.

Clearly, (\ref{Gflux}) and (\ref{nofl1}) are equivalent equations,
so one may choose to work with either of them. In (\ref{nofl1})
the kappa symmetry projector is simpler, but the spinor more complicated.
In the main text we chose to work with (\ref{Gflux}).

\end{document}